\documentclass[twocolumn,epsfig,color]{mnras}
\usepackage{hyperref}
\usepackage[usenames]{color}
\hypersetup{dvips, colorlinks=true, linkcolor=blue, citecolor=blue, filecolor=blue, urlcolor=blue}
\usepackage[dvips]{graphicx}
\usepackage{amssymb}
\usepackage{pifont}
\usepackage{float}

\usepackage[T1]{fontenc}
\usepackage{ae,aecompl}
\usepackage{times}

\newcommand{\mincir}{\raise
-2.truept\hbox{\rlap{\hbox{$\sim$}}\raise5.truept
\hbox{$<$}\ }}
\newcommand{\magcir}{\raise
-2.truept\hbox{\rlap{\hbox{$\sim$}}\raise5.truept
\hbox{$>$}\ }}
\newcommand{\minmag}{\raise-2.truept\hbox{\rlap{\hbox{$<$}}\raise
6.truept\hbox{$>$}\ }}

\newcommand{\hagn}{\mbox{{\sc \small Horizon-AGN}}}

\newcommand{\hdm}{\mbox{{\sc \small Horizon-DM}}}
\newcommand{\hagnn}{\mbox{{\sc \small Horizon-AGN\,\,}}}
\newcommand{\hnoagnn}{\mbox{{\sc \small Horizon-noAGN\,\,}}}
\newcommand{\hdmm}{\mbox{{\sc \small Horizon-DM\,\,}}}

\def\simlt{\lower.5ex\hbox{$\; \buildrel < \over \sim \;$}}
\def\simgt{\lower.5ex\hbox{$\; \buildrel > \over \sim \;$}}
\def\simpropto{\lower.2ex\hbox{$\; \buildrel \propto \over \sim \;$}}

\title[Cusp-core transformations induced by AGN feedback]
{Density profile of dark matter haloes and galaxies in the Horizon-AGN
simulation: the impact of AGN feedback}

\author[S.~Peirani et al.]
 {S\'ebastien Peirani$^{1,2,3,4}$\thanks{E-mail: sebastien.peirani@oca.eu},
 Yohan Dubois$^{2}$,
 Marta Volonteri$^{2}$,
 Julien Devriendt$^{5,6}$,
\newauthor
 Kevin Bundy$^{3}$,
 Joe Silk$^{2,5,7}$,
Christophe Pichon$^{2,8}$,
 Sugata Kaviraj$^{9}$,
 Rapha\"el Gavazzi$^{2}$
\newauthor
and M\'elanie Habouzit$^{2,10}$ 
\\
$^{1}$ Universit\'e C\^ote d'Azur, Observatoire de la C\^ote d'Azur, CNRS, Laboratoire Lagrange, France \\
$^{2}$ Institut d'Astrophysique de Paris (UMR 7095: CNRS \& UPMC), 98 bis Bd Arago, 75014 Paris, France \\
$^{3}$ Kavli IPMU (WPI), UTIAS, The University of Tokyo, Kashiwa, Chiba 277-8583, Japan \\
$^{4}$ Department of Physics, The University of Tokyo, Tokyo 113-0033, Japan\\
$^{5}$ Sub-department of Astrophysics, University of Oxford, Keble Road, Oxford OX1 3RH \\
$^{6}$ Observatoire de Lyon, UMR 5574, 9 avenue Charles Andr\'e, Saint Genis Laval 69561, France\\
$^{7}$ Department of Physics and Astronomy, The Johns Hopkins University Homewood Campus, Baltimore, MD 21218, USA\\
$^{8}$ Korea Institute of Advanced Studies (KIAS) 85 Hoegiro, Dongdaemun-gu, Seoul, 02455, Republic of Korea\\
$^{9}$ Centre for Astrophysics Research, University of Hertfordshire, College Lane, Hatfield, Herts, AL10 9AB, UK\\
$^{10}$ Center for Computational Astrophysics CCA, 160 5th Ave New York\\
 \\
}

\begin{document}

\maketitle

\begin{abstract}

 Using a suite of three large cosmological hydrodynamical simulations, \hagn, \hnoagnn (no AGN
feedback) and \hdmm (no baryons), we investigate how a typical sub-grid model for AGN feedback
 affects the evolution of the inner density profiles of massive dark matter
  haloes and galaxies. Based on direct object-to-object comparisons, we
find that the integrated inner mass and
  density slope differences between objects formed in these three simulations (hereafter, H$_{\rm AGN}$,
H$_{\rm noAGN}$ and H$_{\rm DM}$) significantly evolve
  with time.  More specifically, at high redshift ($z\sim 5$), the
  mean central density profiles of H$_{\rm AGN}$ and
H$_{\rm noAGN}$ dark matter haloes tend to be much steeper
  than their H$_{\rm DM}$ counterparts owing to the rapidly growing baryonic component and ensuing
adiabatic contraction.  By
$z\sim 1.5$, these mean halo density profiles in H$_{\rm AGN}$ 
 have flattened, pummelled by powerful AGN activity (``quasar
  mode''): the integrated inner mass difference gaps with H$_{\rm
    noAGN}$ haloes have widened, and those with H$_{\rm DM}$ haloes
  have narrowed.  Fast forward 9.5 billion years, down to $z=0$, and
the trend reverses: H$_{\rm AGN}$ halo mean density profiles 
drift back to a more cusped shape as AGN feedback efficiency dwindles (``radio mode''),
and the gaps in integrated central mass difference with H$_{\rm noAGN}$
and H$_{\rm DM}$ close and broaden respectively.  On the galaxy side,
the story differs noticeably.
Averaged stellar profile central densities and inner slopes are monotonically
reduced by AGN activity as a function of cosmic time, resulting in better
agreement with local observations.

\end{abstract}


\section{Introduction}





The inner structure of dark matter haloes represents one of the most important
constraint on cosmological and galaxy formation models.
Within the cold dark matter (CDM) paradigm,
earlier works ignoring the effects of gas dissipation and star formation processes,
have suggested that dark matter haloes have spherically-averaged density profiles
that can be well fitted by simple analytical function such as the NFW profile
(Navarro et al. 1996a; Navarro et al. 1997) 
in which the inner slope tends to -1  or
could even be as steep as -1.5 (e.g. Moore et al. 1998; Jing \& Suto 2000).
Later numerical works
favour the Einasto profile with a finite central density
though this profile is still cuspy and close to the NFW one
(Merritt et al. 2006; Stadel et al. 2009; Navarro et al. 2010).

The prediction of a peaked inner density profile
seems to be seriously challenged by observations. In particular,
  dwarfs and low surface-brightness galaxies suggest a much shallower
profile with a relatively flat slope ($\alpha \geq -0.5$ with $\rho(r)\sim r^\alpha$)
(Palunas \& Williams 2000; Salucci \& Burkert 2000; de Blok et al. 2001; Swaters et al. 2003;
Gentile et al. 2004; Spekkens et al. 2005; Goerdt et al. 2006; Walter et al. 2008;
 de Blok et al. 2008; Oh et al. 2011; Walker \& Pe{\~n}arrubia 2011).
On the other extreme of the halo mass scale, galaxy clusters tend to
have a central cusp, as evidenced by  studies  combining strong and weak lensing (e.g., Limousin et al. 2007;
Leonard  et  al.  2007;  Umetsu  et  al.  2007), 
but  shallower slopes than the NFW one  can  also be found  within
the inner $\approx 5$ kpc (Sand et al. 2004; Sand et al. 2008; Newman et al. 2009; 
Newman et al. 2011; Newman et al. 2013; Richtler et al. 2011).
Note that a recent study found a cusp slope over $5-350$ kpc of $\alpha = -1.62$, again from
combined weak and strong lensing analysis of the complete CLASH cluster sample (Zitrin et al. 2015).
At the intermediate halo mass scales, it is worth mentioning that Oldham \& Auger (2016)
have recently demonstrated the presence of a core at the centre of the dark halo of the massive elliptical galaxy M87,
by combining 
stellar kinematics in the central regions with the dynamics of 612 globular clusters.

This so-called ``cusp-core problem'' could imply that the CDM paradigm needs to be revised 
 to account for dark matter self-interaction 
(Yoshida et al. 2000; Burkert 2000; Kochanek \& White 2000;  Spergel \& Steinhardt 2000;
Dav{\'e} et al. 2001; Ahn \& Shapiro 2005; Vogelsberger et al. 2014b, Elbert et al. 2015; Lin \& Loeb 2016),
a warmer dark matter candidate (Col{\'{\i}}n et al. 2000;
Bode et al. 2001, Lovell et al. 2012) or
an ultralight axion-particle (e.g., Marsh \& Pop 2015), and/or 
a more exotic coupling between dark matter and dark energy (e.g.
Macci\`o et al. 2015). Alternatively, it could simply mean that
baryonic processes play a key role in shaping the inner parts
of halos and galaxies. Indeed, several astrophysical solutions have been proposed to reconcile observations with theoretical predictions.
Stellar feedback could produce rapid variations of the
gravitational potential through substantial gas mass outflows from the
central region. This would flatten the inner density profile of the
dark matter halo (Navarro et al. 1996b; Gnedin \& Zhao 2002; Read \& Gilmore 2005;
Mashchenko et al. 2006, 2008; Ogiya \& Mori 2011, 2014;
Governato et al. 2012; Pontzen \& Governato 2012; Macci{\`o} et al. 2012; Teyssier et al. 2013;
 O{\~n}orbe et al. 2015; Chan et al. 2015; El-Zant et al. 2016; Del Popolo \& Pace 2016).
On the other hand, dark matter can also be gravitationally ``heated'' by baryons
through dynamical friction caused either by self-gravitating gas clouds orbiting near the center of the galaxy
(El-Zant et al. 2001, El-Zant et al. 2004; Jardel \& Sellwood 2009;
Lackner \& Ostriker 2010; Cole et al. 2011, Del Popolo \& Pace 2016)
by the presence of a stellar bar (Weinberg \& Katz 2002;
Holley-Bockelmann et al. 2005; Sellwood 2008),
by the radiation recoil from coalescing black holes (Merritt et al 2004), 
or by processes which transfer of angular momentum from baryonic to dark matter (Tonini et al. 2006, Del Popolo 2009, 2012, 2014).

On larger mass scales, numerical simulations from Peirani et al. (2008)
(see also Duffy et al. 2010; Dubois et al. 2010; Teyssier et al. 2011; Martizzi et al. 2012, 2013, Ragone-Figueroa \& Granato 2011;  Ragone-Figueroa et al. 2012, 2013)
have argued that active galactic nuclei (AGN) feedback plays a similar
role to that of stellar feedback in smaller systems: it can heat/expel 
large amounts of gas from the central regions of galaxy groups and
clusters. A fraction of this gas then cools and returns to the centre,
generating repeated cycles of significant inflows/outflows which in turn
cause rapid fluctuations of the gravitational potential,
steepening/flattening out the inner dark matter halo and galaxy stellar density profiles. 
Such a mechanism is commonly invoked to explain the substantial body of 
observational evidence that the majority of massive elliptical
or cD galaxies exhibit very shallow slopes in their inner ($\approx 1$
kpc) stellar surface brightness profiles (Kormendy 1999; Quillen, Bower \& Stritzinger 2000; Laine et al. 2003; Graham 2004; Trujillo et al. 2004;
Lauer et al. 2005; Ferrarese et al. 2006; C\^ot\'e et al. 2007; Kormendy et al. 2009; Graham 2013). 
A related phenomenon is the formation of cores  within the central 100 pc of massive
ellipticals. These are believed to be formed dynamically, by the
scouring effect of possibly stalled supermassive black hole (SMBH) binaries  
at $\sim 10$ pc separation (see for instance Faber et al. 1997; Thomas et
al. 2014 and references therein).  Larger cores are also found, for more widely separated SMBH pairs,  of
up to $\sim 500$ pc extent (e.g. Mazzalay et al. 2016 and references therein).

In the present paper, we aim to extend previous theoretical work on the role
played by AGN feedback, using a statistically representative sample
spanning a comprehensive range of dark matter halo and galaxies masses
and looking at the evolution of their inner density profiles throughout a
considerable fraction of the age of the Universe.
This sample is extracted from our state-of-the-art hydrodynamical cosmological simulation
 \hagnn (Dubois et al. 2014) which includes gas cooling, star formation, stellar and AGN feedback,
 and that we compare with two other simulations \hnoagnn (no AGN feedback) and \hdmm (no baryons) 
stripped down in terms of modelled physical processes but featuring
identical initial consitions. These simulations have been used to
highlight the role of AGN feedback (Volonteri et al. 2016) in
regulating the baryon content of massive galaxies (Kaviraj et
al. 2017; Beckmann et al 2016) and their
 morphological transformations (Welker et al. 2017; Dubois et al. 2016).

The paper is organized as follows. Section 2
briefly summarises the numerical modelling upon which this work is based (simulations and post-processing).
Section 3 and 4 present our main results
concerning the evolution of the inner density profiles of massive dark matter haloes and galaxies respectively. 
Finally, we put forward and discuss our conclusions in Section 5.

\section{Numerical modelling}

\subsection{The three simulations  H$_{\rm AGN}$, H$_{\rm noAGN}$, and H$_{\rm DM}$ }

In this paper, we analyse and compare two large cosmological hydrodynamical simulations, 
\hagnn (H$_{\rm AGN}$),
 \hnoagnn (H$_{\rm noAGN}$) and one dark matter only cosmological simulation \hdmm (H$_{\rm DM})$.
\hagnn is already described in Dubois et al. (2014), so we only summarise here its main features.
We adopt a standard $\Lambda$CDM cosmology with total matter density $\Omega_{\rm m}=0.272$, 
dark energy density $\Omega_\Lambda=0.728$, amplitude of the matter power spectrum $\sigma_8=0.81$,
 baryon density $\Omega_{\rm b}=0.045$, Hubble constant H$_0=70.4 \, \rm km\,s^{-1}\,Mpc^{-1}$, and $n_s=0.967$ 
compatible with the WMAP-7 .
The size of the simulated volume is $L_{\rm box}=100 \, h^{-1}\rm\,Mpc$ on a side, and it contains
 $1024^3$ dark matter (DM) particles, which results in a DM mass resolution of
 $M_{\rm DM, res}=8.27\times 10^7 \, \rm {\rm M}_\odot$.
The simulation is run with the {\sc ramses} code (Teyssier 2002), and the
 initially uniform grid is adaptively refined down to $\Delta x=1$ proper kpc at all times. 
Refinement is triggered in a quasi-Lagrangian manner: if the number of DM particles
 becomes greater than 8, or the total baryonic mass reaches 8 times the initial DM mass
 resolution in a cell.

Gas can radiatively cool down to $10^4\, \rm K$ through H and He collisions with a
 contribution from metals using rates tabulated by Sutherland \& Dopita (1993). 
Heating from a uniform UV background takes place after redshift $z_{\rm reion} = 10$
 following Haardt \& Madau (1996). 
The star formation process is modelled using a Schmidt law: $\dot \rho_*= \epsilon_* {\rho / t_{\rm ff}}$ for
 gas number density above $n_0=0.1\, \rm H\, cm^{-3}$,  where $\dot \rho_*$ is the star formation rate density,
 $\epsilon_*=0.02$ the constant star formation efficiency, and $t_{\rm ff}$ the local free-fall time of the gas.
The stellar mass resolution is $M_{\rm *,res}=2\times 10^6 \,\rm M_\odot$.
Feedback from stellar winds, supernovae type Ia and type II are also taken into account for mass,
 energy and metal release (Kimm, 2012).

Black hole (BH) formation is also included, and BHs accrete gas at a Bondi-capped-at-Eddington 
rate and coalesce when they form a tight enough binary.
They also release energy in a quasar (heating) or radio (kinetic jet) mode when the accretion rate is
 above (below) one per cent of Eddington, with efficiencies tuned to match the BH-galaxy
 scaling relations (see Dubois et al. 2012 for detail).
The presence of both quasar and radio modes is supported by recent observations. In particular, using MaNGA data
 (Bundy et al. 2015), Cheung et al. (2016) report the presence of 
 bi-symmetric emission features
in the centre of quiescent galaxies of mass around $ 2\times 10^{10} {\rm M}_\odot$
from which they infer the presence of centrally driven winds.
On top of the fact that such ``red geysers'' galaxies seem to be very
common at this mass scale (Bundy et al., in prep), the  
energy released by their SMBHs is capable of driving the observed winds and displays
a mechanical content sufficient to suppress star formation. 
It is therefore very likely that such kinetic winds (radio mode) play a crucial role in galaxy formation 
and should be taken into account in numerical models (see for instance, Weinberger et al. 2017).

\hnoagnn and  \hdmm were performed using the same set of initial
conditions and sub-grid modelling of physical processes but
 with no BH formation (and therefore no AGN feedback) and baryons respectively.

\subsection{Dark matter halo and galaxy catalogues }

Dark matter haloes are identified using the \mbox{{\sc \small
    AdaptaHOP}} (sub)halo finder (Aubert et al. 2004, Tweed et al. 2009).
In our different catalogues, host haloes and subhaloes are studied separately.
 Since we are particularly interested in studying the very inner part
 of dark matter haloes, a robust definition of their centre is critical. In general, the position of
the most bound particle yields an accurate estimate, especially in the
case of \hnoagnn and \hdmm as haloes in these simulations have cuspy
inner profiles. However, as far as \hagnn is concerned, as 
we will see in the remainder of this paper, the dark matter profiles
of haloes hosting large galaxies can be flatter. Therefore, for some (rare) object in this latter simulation, the centre can be associated
with a substructure which is offset from the ``real'' centre of
mass and could lead to the attribution of a spurious core. To
circumvent this issue, we use a shrinking sphere approach (Power et
al. 2003) whereby starting from the virial radius, we
recursively identify the centre of mass within spheres 10\% smaller in
linear size at each iteration. We stop the procedure once the sphere
reaches a 2 kpc radius and identify the centre of the halo with its
densest particle. Twenty neighbours are used to compute the local
density. Only structures with an average density larger than 200 times the average matter
density and containing more than 100 particles become part of the (sub)halo
catalogue.

\begin{figure}
\rotatebox{0}{\includegraphics[width=\columnwidth]{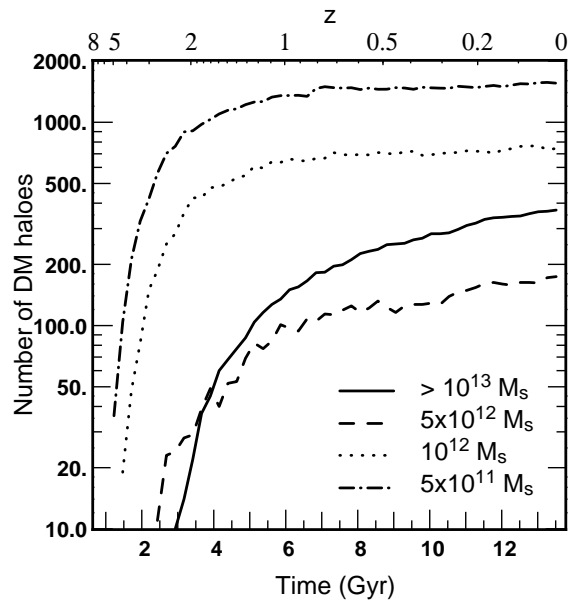}}
\caption{The number of dark matter haloes matched between the three simulations for
our four distinct fixed mass intervals namely $\sim 5\times10^{11}  {\rm M}_\odot$ (dotted-dashed line), $\sim 10^{12} {\rm M}_\odot$ (dotted line),
 $\sim 5\times10^{12}  {\rm M}_\odot$ (dashed line) and
$\geq 10^{13} {\rm M}_\odot$ (solid line). We derived statistics only when 10 objects can be
identified at a specific redshift.}
\label{fig0}
 \end{figure}

Galaxies are also identified with \mbox{{\sc \small AdaptaHOP}} but applied to star particles.
In this case, we use the most bound particle as the definition of the centre of mass and only galaxies
with more than 50 star particles are part of the final catalogue.

Finally, we build the merger trees of all the dark matter haloes and galaxies for each simulation, using 
\mbox{{\sc \small TreeMaker}} (Tweed et al. 2009) to link 52 outputs equally
spaced in time between $z\sim5.8$ and $z=0$, i.e. with a time resolution
of $\sim$ 250 Myr.

These procedures yield, for the  \hagnn run at redshift zero, about $250000$ haloes and subhaloes
 with a mass greater than $10^{10} {\rm M}_\odot$ and $125000$ galaxies  with a mass greater than
 $10^{8} {\rm M}_\odot$.

\subsection{Matching dark matter haloes and galaxies}

Since we start from the same initial conditions, each dark matter particle
 possesses an identity which is identical in any of the 3
 simulations. Thus, if 75\% or more of the particles of any given halo
 in the \hagnn run also belong to a halo identified in the \hnoagnn or
 \hdmm runs, we initially assume that these haloes are twins. However,
 if the mass ratio of the matching pair is greater than 10 (or lower than 0.1), we exclude it 
from our comparison sample. This last step is rendered necessary
because sub-structures can get their particles stripped
 by the host halo at different times and with different intensities 
in the three simulations. As a result, a subhalo could become twinned
with a much more massive host halo if its equivalent subhalo in the other simulation has already
been destroyed (or has become too small to be detected).
In general, we are able to match more than 85\% of dark matter objects
at any redshift by applying these two criteria.

We cannot implement the same procedure for galaxies since a given stellar
particle is not necessarily created at the exact same moment in the
very same galaxy in \hagnn and \hnoagnn. Moreover, the total number of
stellar particles will also differ as it depends on the impact AGN
feedback has on the star formation process. 
Therefore, instead of relying upon a common identity of the stellar
particles they are composed of to directly match objects between runs, we first couple each galaxy to a 
host dark matter halo in their parent simulation. We determine these
galaxy-halo pairs by picking the most massive galaxy whose centre is
located within a sphere of radius equal to $5\%$ of the virial radius
of its host halo. Galaxy twins between runs are then determined through
the matching of their host halo as previously described. 

To illustrate the typical efficiency of such a procedure, at $z=0$, 
we are able to match about $68\%$  of H$_{\rm AGN}$ galaxies   
with a mass greater $10^{10} {\rm M}_\odot$ to H$_{\rm noAGN}$ counterparts.
This fraction might seem a bit low but our matching algorithm requires
three steps to establish the link (galaxy to host halo, host halo to host halo twin, host
halo twin to galaxy twin), with a number of objects  dropping  out  of  the  sample  at  each  step.
Moreover, identifying a  galaxy  with  its  host  halo  can  be  challenging,  especially
in  dense  environments where  interactions are more common. So in order to avoid complex 
situations especially during merger processes, we chose restrictive parameters. 
In particular, we picked the most massive galaxy whose centre is located within a sphere
 of radius equal to 5\% of the virial radius. 
Increasing this latter value to 10\% for instance would increase the number
 of matched galaxies (see Chisari et al. 2017). However, 
during very complex merger processes especially in dense environment where galaxy centre
 can have a big offset with respect to the centre of its host halo, it is not guaranteed to 
select the good galaxy.
Nevertheless, we have checked that relaxing the quite stringent criteria adopted in
this work  improves the matching fraction, but since it also
increases the number of false matches and does not alter any of our
conclusions, we prefer to restrict ourselves to the more conservative
sample defined in this section.

\section{Dark matter halo density  profiles }
Let us first study the evolution of the inner density profiles of dark matter haloes in the three simulations.
\subsection{Definitions}

\begin{figure*}
\rotatebox{0}{\includegraphics[width=16.5cm]{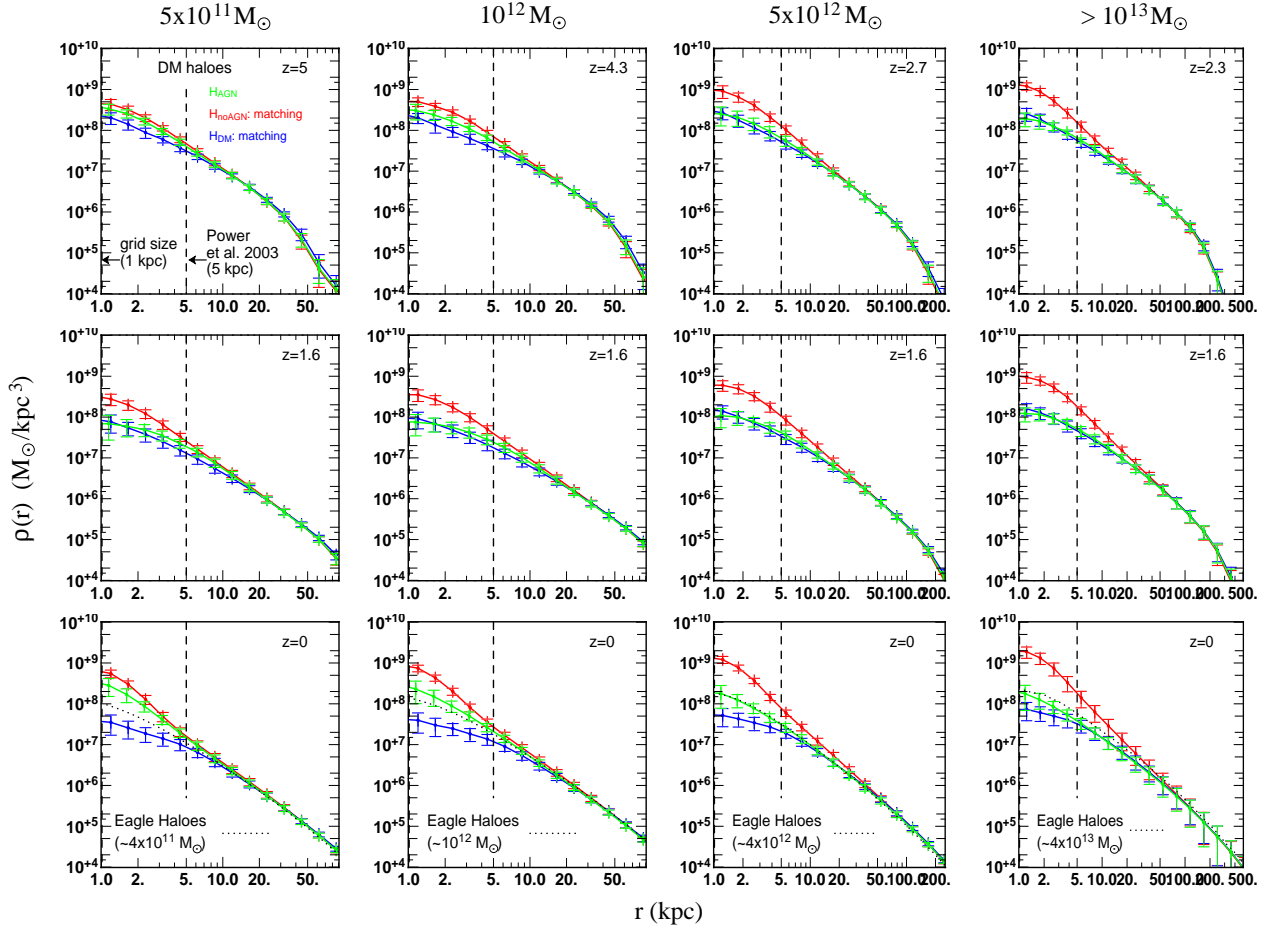}}
\caption{The evolution of the mean density profiles of dark matter haloes
 extracted  from H$_{\rm AGN}$ (green lines),
 H$_{\rm noAGN}$ (red lines) and  H$_{\rm DM}$ (blue lines). We focus on four distinct fixed
mass intervals: $\sim 5\times10^{11}  {\rm M}_\odot$ (first column), $\sim 10^{12} {\rm M}_\odot$ (second column),
 $\sim 5\times10^{12}  {\rm M}_\odot$ (third column) and
$\geq 10^{13} {\rm M}_\odot$ (fourth column). Three different epochs are also considered:
 high redshift (first line),
intermediate redshift (second line) and $z=0$ (third line). For indicative purposes only, the two vertical dashed lines
at $r=1$ kpc and $r=5$ kpc represent respectively the simulation grid size and a recommended
 resolution limit following Power et al. (2003). The error bars correspond to the dispersion.
 These plots suggest that H$_{\rm noAGN}$ haloes have always very dense and cuspy
central regions. On the contrary, AGN feedback tends to flatten the profiles especially
at intermediate redshifts ($z\sim [1.6-2.7]$) whereas a ``cusp regeneration'' is observed at $z=0$.
Finally, we also show at $z=0$ the mean density profiles of dark matter haloes of similar mass range
extracted from the Eagle simulation (Schaller et al. 2015a). The latter results suggest 
there are some slight differences especially for lower mass halos ($5\times10^{11} M_\odot$) but quite consistent
results for massive ones.}
\label{fig1}
 \end{figure*}

Our investigation starts with the evolution of the dark matter component.
The questions we want to address here are twofold: 1) Does AGN feedback  noticeably modify the  
inner density profiles of dark matter haloes? and 2) How does the 
difference between the inner density profiles of H$_{\rm AGN}$ and
H$_{\rm noAGN}$ (or H$_{\rm DM}$) haloes evolve?
To do so, we use a systematic 
object-to-object comparison  between the \hagn, \hnoagnn and \hdmm
simulations based on the matching procedure described in the previous section.
Note that when calculating the density profiles of haloes
in the H$_{\rm DM}$ run, we rescale the mass of the DM particles by a factor 
$(\Omega_{\rm m} -\Omega_{\rm b})/\Omega_{\rm m}$ to make it
identical to the mass of DM particles in the baryonic runs.

\begin{figure}
\rotatebox{0}{\includegraphics[width=8.2cm]{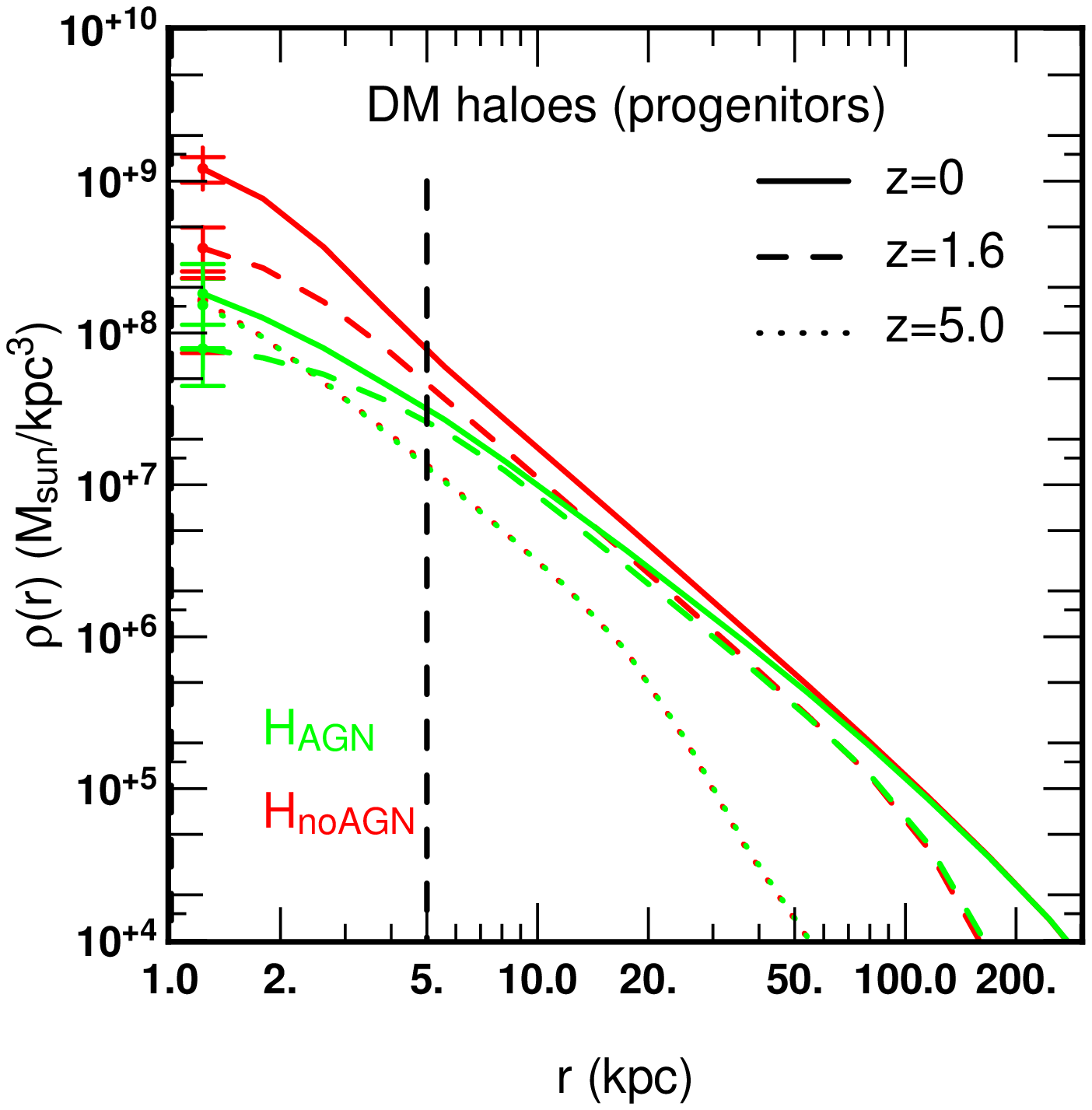}}
\rotatebox{0}{\includegraphics[width=8.2cm]{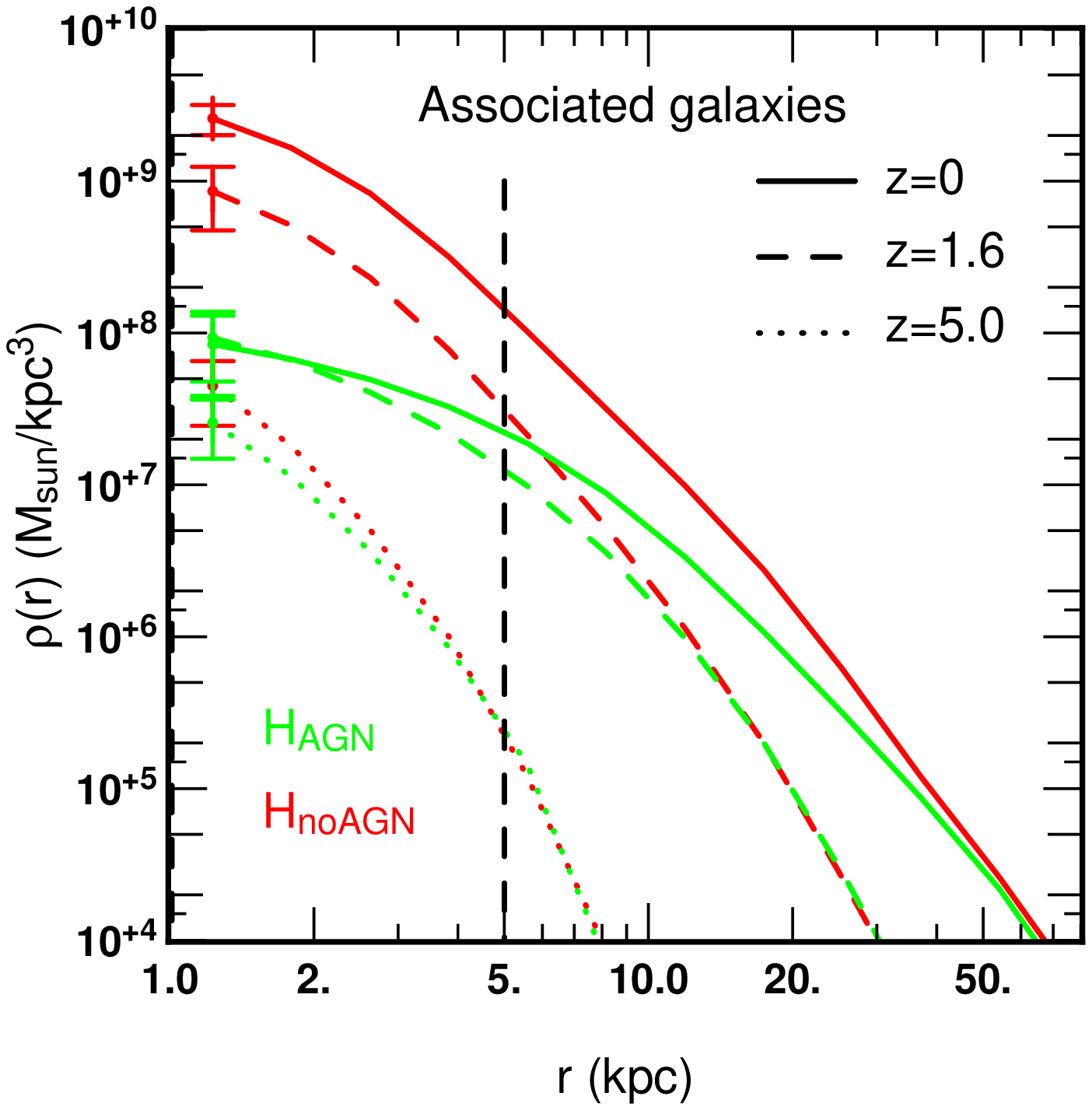}}
\caption{Upper panel: the evolution of the mean density profiles of the progenitors of
H$_{\rm AGN}$ dark matter haloes of mass $\sim 5\times10^{12}  {\rm M}_\odot$  at $z=0$ (green lines).
Three different redshifts have been considered: $z=5$ (dotted line), $z=1.6$  (dashed line)
and $z=0$ (solid line). Results for the H$_{\rm noAGN}$ counterpart profiles are shown in 
red colors. For indicative purposes only, the two vertical dashed lines
at $r=1$ kpc and $r=5$ kpc represent respectively the simulation grid size and a
recommended lower resolution limit following Power et al. (2003).
For clarity, we do not show either the  H$_{\rm DM}$ counterpart profiles
or all the  dispersion (which are similar to those in Fig. \ref{fig1}).
As for Fig. \ref{fig1}, the mean density profile of dark matter haloes is flatter at intermediate
redshift ($z\sim 1.6$) and steeper at high and low z. 
The lower panel shows the corresponding situation for the associated galaxies.
In this case, when AGN feedback is included, the profiles of galaxies progressively
 flatten all the way to $z=0$. }
\label{fig2}
 \end{figure}

In the following, we split our H$_{\rm AGN}$ DM halo sample into four different mass intervals: 
$5\times10^{11} (\pm 10\%) {\rm M}_\odot$,
$10^{12}  (\pm 10\%) {\rm M}_\odot$,
$5\times10^{12}  (\pm 10\%) {\rm M}_\odot$ and
$\geq 10^{13} {\rm M}_\odot$, which we match to their H$_{\rm noAGN}$ and H$_{\rm DM}$ counterparts.
For each of these four mass bins, we then compute the mean density profiles
(binned in spherical shells equally spaced in $\log r$), i.e.  
$\rho_{\rm AGN}(r)$, $\rho_{\rm noAGN}(r)$ and $\rho_{\rm DM}(r)$ at every redshift. This allows
us to consider the evolution of density profiles at fixed halo mass.
In addition, we have also considered
the evolution of the density profiles of the progenitors of halos
within these mass bins at $z=0$. 
Thanks to our large simulated volume,
each sub-sample consists in general of ten objects or more in the most massive bin and thousands of objects
in the least massive one. However, when this is not the case (at high
redshift for the most massive objects: see Fig. \ref{fig0} or  Fig. \ref{fig1}),
 we lower the redshift until a
minimum of ten haloes of that mass can be identified and
an average density profile computed.

\begin{figure}
\rotatebox{0}{\includegraphics[width=\columnwidth]{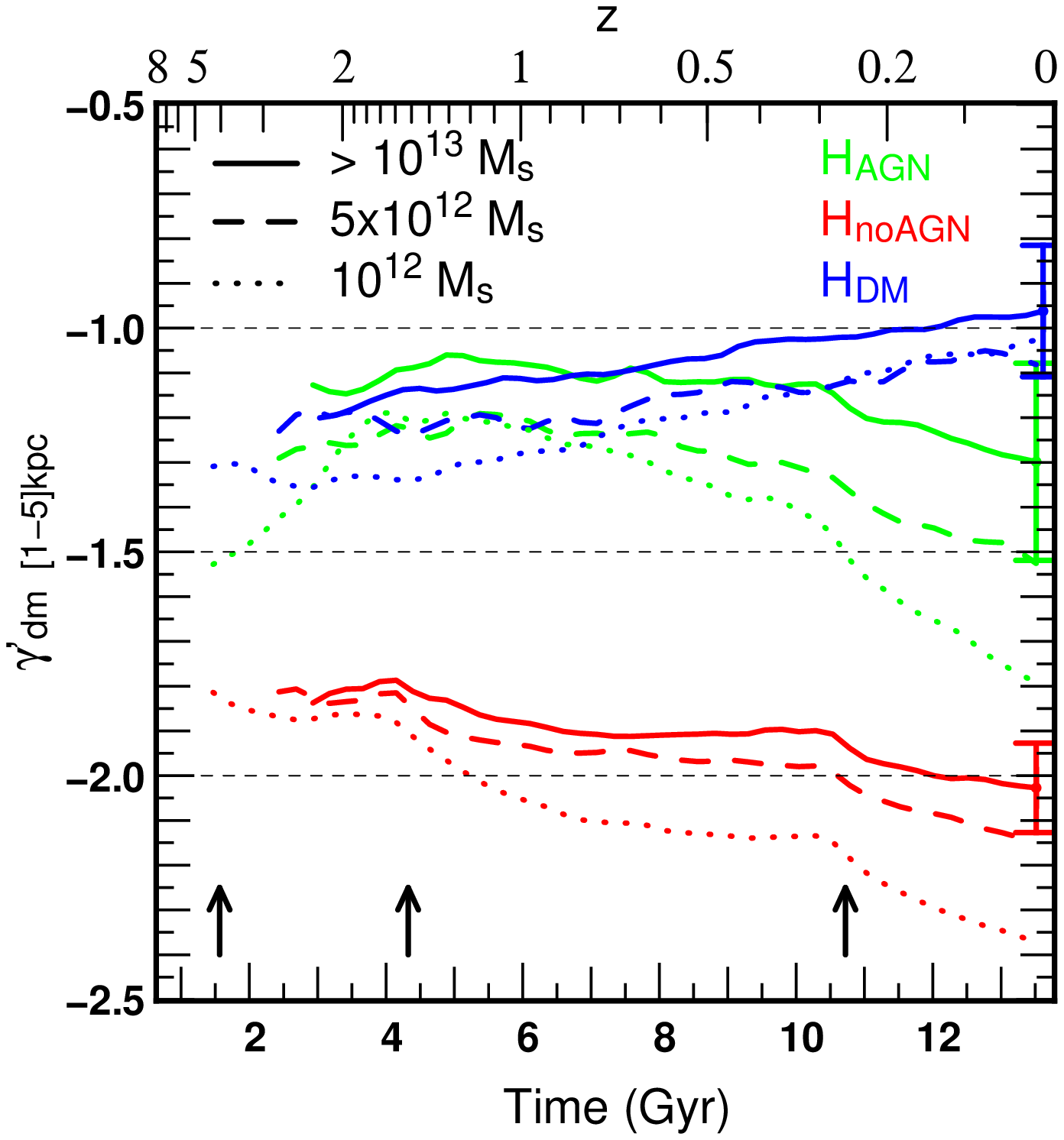}}
\rotatebox{0}{\includegraphics[width=\columnwidth]{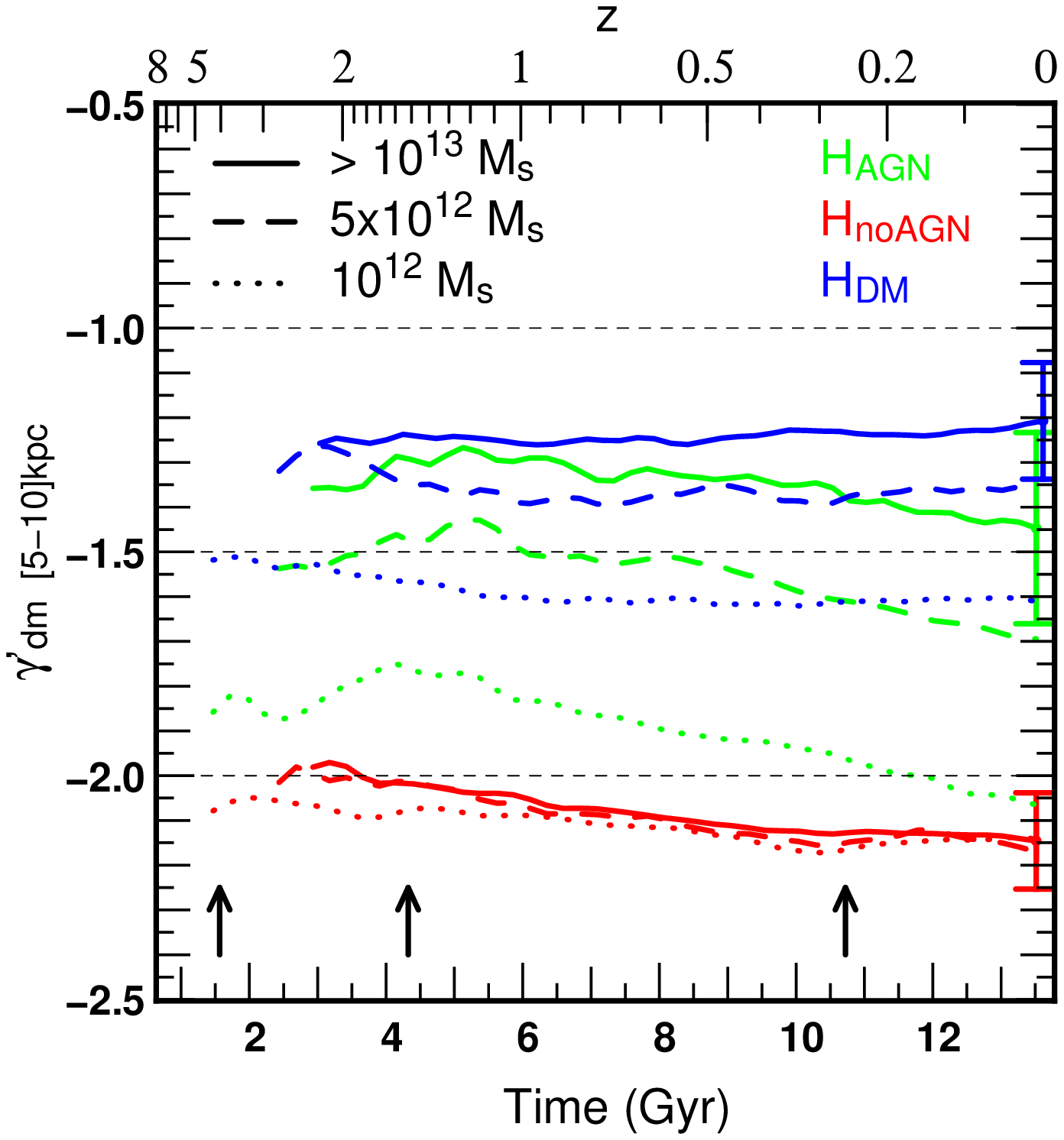}}
\caption{Time evolution of the mean mass-weighted density slope $\gamma'_{\rm dm}$ estimated 
either within  [1-5] kpc (upper panel) or  [5-10] kpc (lower panel) of DM haloes with masses $10^{12} {\rm M}_\odot$ (dotted lines), $5\times 10^{12} {\rm M}_\odot$
(dashed lines)  and $> 10^{13} {\rm M}_\odot$ (solid lines). 
Slopes measured for \hagn haloes and matched \hnoagnn and \hdmm counterparts are coloured in green, red and blue respectively. 
The three arrows indicate the times when a new refinement level is
added in the simulations.
For clarity we do not  show the corresponding evolution for the $5\times 10^{11} {\rm M}_\odot$ mass
interval  as it is similar to those derived for haloes of mass $10^{12} {\rm M}_\odot$. Typical 
dispersions are indicated by vertical error bars at $z=0$.
}
\label{fig3}
 \end{figure}

\begin{figure}
\rotatebox{0}{\includegraphics[width=\columnwidth]{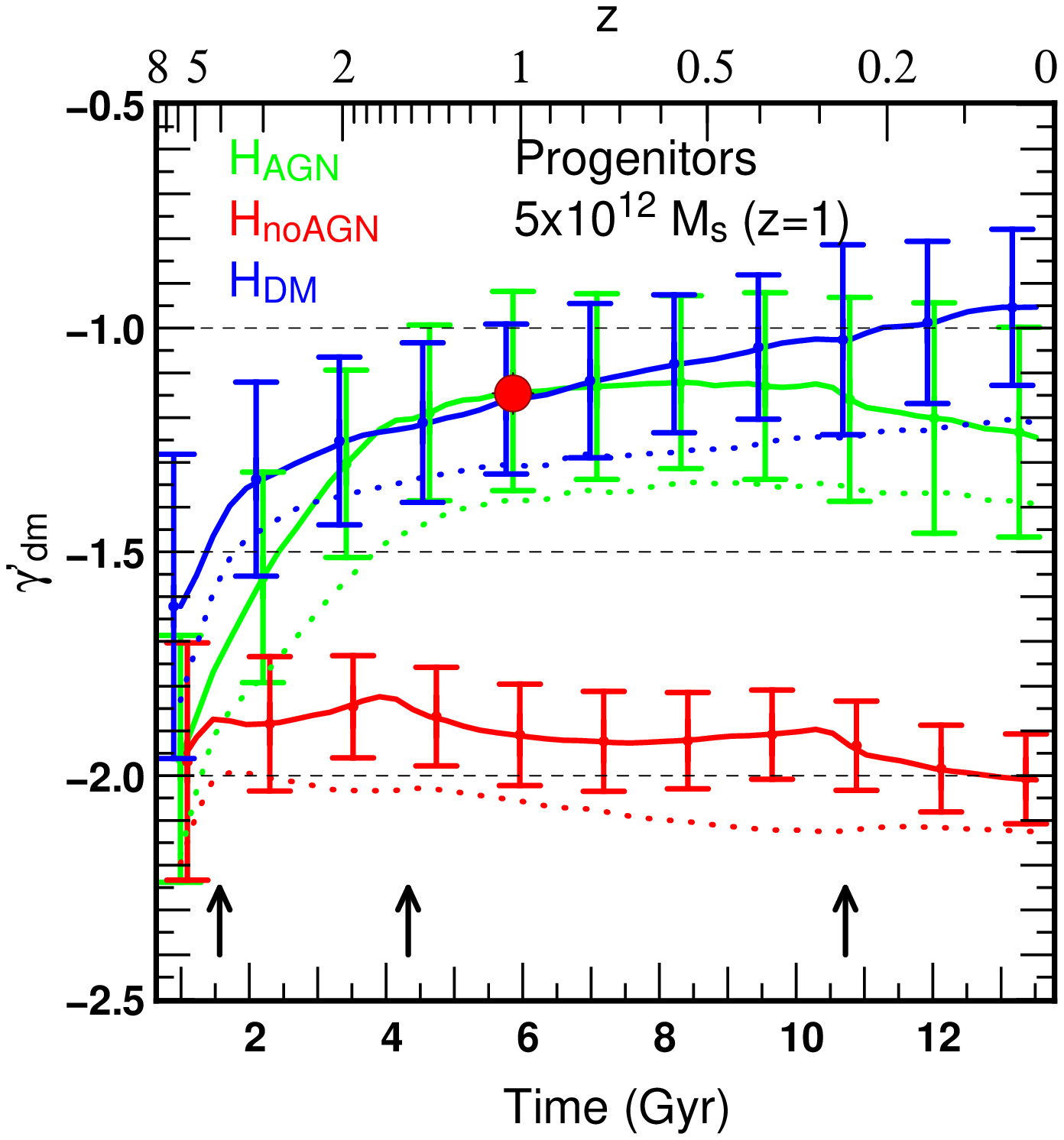}}
\rotatebox{0}{\includegraphics[width=\columnwidth]{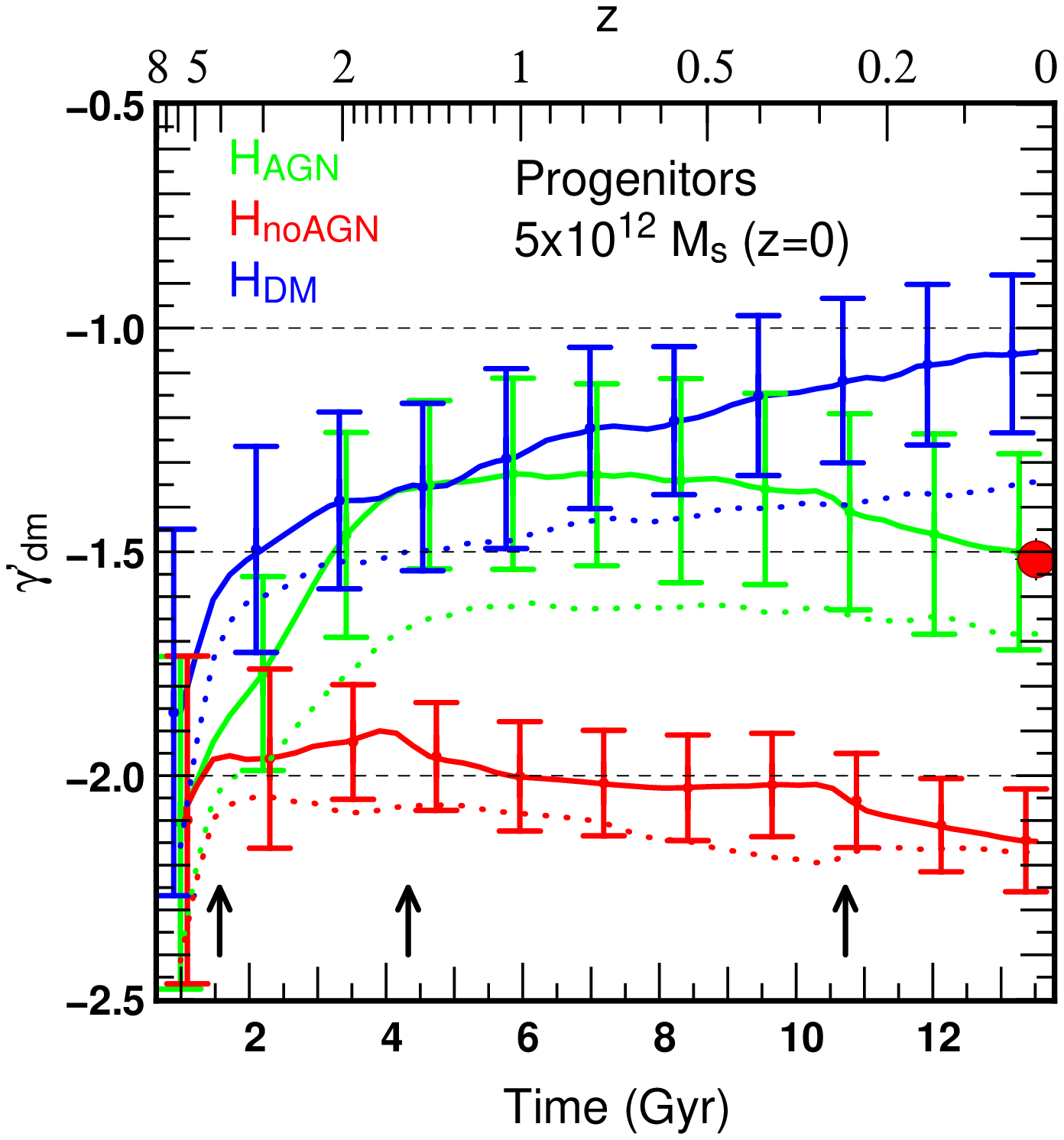}}
\caption{Time evolution of the mass-weighted density slope $\gamma'_{\rm dm}$ 
of progenitors of DM haloes with a mass of 
$5\times 10^{12} {\rm M}_\odot$ at $z=1$ (upper panel) and $z=0.0$ (lower panel).
 The mean slope has been estimated either in the range of [1-5] kpc (solid lines) or [5-10] kpc
 (dotted lines). Results
from \hagn, \hnoagnn and \hdmm simulations are coloured in green, red and blue respectively.
The three arrows indicate the times when a new refinement level is
added in the simulations.
The error bars  correspond to the 
standard deviations. To facilitate comparison, filled red circles in
each panel indicate the values of $\gamma'_{\rm dm}$ when  H$_{\rm AGN}$ dark matter haloes reach a
mass of $5\times 10^{12} {\rm M}_\odot$. In spite of belonging to the
same mass sub-sample, H$_{\rm AGN}$  haloes at  $z=0.0$ 
tend to have steeper profiles than those at $z=1$.}
\label{fig4}
 \end{figure}

In order to  study the (relative) evolutions of the mean inner density profiles
of dark matter haloes extracted from the three different simulations, we use two estimators.
The first one is the  {\it mass-weighted density slope within $r_1$ and $r_2$} 
introduced by Dutton \& Treu (2014): 
\begin{equation}
\gamma' = \frac{1}{M(r)}\int_{r_1}^{r_2}\gamma(x)4\pi r^2 \rho(x)\, \mathrm{d}x\,,
\label{equ1}
\end{equation}
\noindent
where $\gamma\equiv {\it d}$\,log\,${\it \rho}$$/{\it d}\,$log$\, {\it r}$ is the
local logarithmic slope  of the density profile $\rho$ and $M$ the local mass.
Using a discrete representation of the density profiles, we measure
$\gamma(r)$ and  $M(r)$ for each spherical shell centered on position
$r$. $\gamma'$ is estimated in the range $[r_1-r_2]=[1.0-5]$ kpc.

Secondly, in order to quantitatively study the evolution of the gap between the different density profiles, 
we compute the quantity $A_{\rm DM}$ and $A_{\rm noAGN}$ defined by:
\begin{equation}
\hskip -0.2cm
A_{\rm DM|noAGN}=\int_{\log r_1}^{\log r_2}\!\!\frac{\rho_{\rm DM|noAGN}(r)-\rho_{\rm AGN}(r)}{\rho_{\rm AGN}(r)} d\log r\,.
\label{equ2}
\end{equation}
 $A_{\rm DM}$ and $A_{\rm noAGN}$ give an estimation of the gap between
 the  mean profiles of H$_{\rm noAGN}$ and H$_{\rm DM}$ haloes with respect to the mean profiles of H$_{\rm AGN}$ haloes.
In the following, $A_{\rm DM}$ and $A_{\rm noAGN}$ have been estimated 
in the range $[r_1-r_2]=[1-10]$ kpc. 

Note that the minimal value adopted for $r_1$ is dictated by our spatial
resolution, which does not allow us to reach smaller scales. Spatial resolution is
also the main reason why we decide not to use a fixed fraction of the virial
radius to define $r_1$ (and $r_2$): we would be restricted to selecting (at least) 
$r_1 = 0.01 \times r_{\mathrm vir}$ at high redshift because haloes are
more compact and we cannot probe their inner kpc. In turn, this would
then translate into $r_1 \geq 10$kpc for galaxy cluster size
haloes at $z=0$, which somewhat contradicts our purpose to examine their inner density
profile to the best of our ability. 
For these reasons, we adopt instead a simple fixed physical scale for
$r_1$ and $r_2$: we probe the inner 1-5 ($\gamma'$) or 1-10 ($A_{\rm DM|noAGN}$) kpc of DM haloes
for any redshift and halo mass.
The different choice of value for $r_2$ used for $\gamma'$ and $A_{\rm  DM|noAGN}$ illustrates the fact that the
results presented in this paper are robust to variations of up to a factor 2 in the values of
$r_1$ and $r_2$.

\subsection{Visual inspection: a three phases scenario?}

Fig. \ref{fig1} shows the averaged spherical density profiles $\rho_{\rm AGN}$, $\rho_{\rm noAGN}$ and 
$\rho_{\rm DM}$ derived for haloes pertaining to our different mass sub-samples and 
at three different redshifts. At a given time, they clearly appear different from one other especially
 in the central region ($r \lesssim 20$ kpc).  Conversely, they seem indistinguishable at large radii,
 suggesting that the presence of baryons and/or AGN feedback  induces
 noticeable effects only on small scales.

Now, if one analyses the evolution of the mean H$_{\rm DM}$, H$_{\rm noAGN}$ and H$_{\rm AGN}$
 halo density profiles separately, 
clear trends can be noticed. First, as expected from numerous previous studies, the H$_{\rm DM}$  density profiles
are  always centrally cuspy: their inner slopes are consistent
 with a NFW profile (as shown in the next section). Note also that at
 fixed halo mass, the H$_{\rm DM}$ 
density profiles are more extended, less concentrated, at present times than at high redshift.
 This well known result is mainly due to the fact that 
low redshift haloes have undergone  more (major) mergers than their
high redshift analogues, and these mergers tend to diffuse material at larger radii
 (see for instance Klypin et al. 2016). As a consequence, for a fixed mass interval,
 the density in the halo inner region progressively decreases as more and more mass  
is distributed at large radii.
The H$_{\rm noAGN}$ halo density profiles are always much steeper and have higher central
values compared to those in the other simulations. For a fixed mass interval,
no significant variation in profile shape as a function of redshift is observed.
Finally,  the H$_{\rm AGN}$ density profiles present  a  more complex evolution. 
At high redshift ($z\geq 4$) 
the halo density profiles of the two first mass bins (i.e. $5\times10^{11} {\rm M}_\odot$  and
$10^{12} {\rm M}_\odot$) are quite steep and close to their  H$_{\rm noAGN}$ matches.
At intermediate redshifts ($z\sim [2.7-1.6]$), 
 they appear to have flattened 
and can even have lower central density values ($\leq 5$ kpc)  than to their H$_{\rm DM}$ counterparts.
Finally, at present times, the density profiles
of H$_{\rm AGN}$ haloes steepen and approach again those of
their H$_{\rm noAGN}$ twins. 
In view of these results, the evolution of H$_{\rm AGN}$ halo density profiles therefore
 seems to follow three distinct phases.

This scenario is confirmed when studying the evolution of the progenitors of dark matter
haloes of mass $5\times10^{12} {\rm M}_\odot$ at $z=0$. The variations of their density profiles
are shown at the same three redshifts in Fig. \ref{fig2}. 
H$_{\rm AGN}$ haloes clearly present a mean density profile that is
shallower at $z=1.6$  than at $z=5$ or $z=0$.
This strongly suggests the existence of successive phases
 of expansion and contraction of the mean inner DM halo density profile.  Note that at $z=5$,  $\rho_{\rm AGN}$ and
$\rho_{\rm noAGN}$ are almost indistinguishable (red and green dotted
lines on Fig \ref{fig2}).
 Since the density profiles of proto-dark matter haloes (i.e. at
very high redshifts) of the three simulations are expected to be identical (as galaxy formation has
not happened yet), this indicates that H$_{\rm AGN}$  and
 H$_{\rm noAGN}$ haloes density profiles have undergone a nigh
 identical first phase of condensation before $z\sim 5$.
It is worth mentioning that similar trends are obtained when following the 
progenitors of the other halo mass sub-samples.
Finally, as we will see in more detail in section \ref{section_galaxies},
the situation differs for the evolution of galaxy profiles, in the sense that
H$_{\rm AGN}$ stellar density profiles remain quite shallow at low redshift.

\subsection{Quantitative trends}

The three redshift snapshots presented in Fig. \ref{fig1} and \ref{fig2} provide
a useful qualitative impression of the general evolution of the mean density profiles of 
H$_{\rm DM}$, H$_{\rm noAGN}$ and H$_{\rm AGN}$ haloes. In order to
derive more quantitative estimates, we now look at the evolution of
the mass-weighted slope of their inner density profile, $\gamma'_{\rm dm}$, 
as defined by equation \ref{equ1}.

Fig. \ref{fig3}  shows  the evolution
 of $\gamma'_{\rm dm}$ for H$_{\rm AGN}$ DM haloes with a fixed mass of $10^{12} {\rm M}_\odot$ (dotted lines), $5\times 10^{12} {\rm M}_\odot$
(dashed lines)  and $> 10^{13} {\rm M}_\odot$ (solid lines). Results for their H$_{\rm noAGN}$ and H$_{\rm DM}$  
counterparts are also presented in the figure.
First, as far as H$_{\rm DM}$ haloes are concerned, values of $\gamma'_{\rm dm}$ are very close to $-1$,
consistent with  NFW  profile expectations, although we note that $\gamma'_{\rm dm}$ is slightly increasing with
 time by about 0.3.  As haloes become more and more extended at low redshift, 
the range of [1-5] kpc is probing a relatively ``deeper'' region in
terms of fraction of the virial radius compared to that probed for
haloes of the same mass at higher redshifts. 
Second,  as expected from adiabatic contraction considerations
(e.g. Blumenthal et al 1986), the inner density profiles of H$_{\rm noAGN}$   are always very steep with slopes that
are close to $-2$. In this case a slight decrease, of a similar
amplitude to the increase noted for H$_{\rm AGN}$ haloes, is observed in the evolution of
$\gamma'_{\rm dm}$.
Finally, H$_{\rm AGN}$ haloes  with a fixed mass of $10^{12} {\rm
  M}_\odot$ clearly exhibit shallower inner density slopes at z$\sim$[1.5-1.0]
than at higher or lower redshift, thus confirming the more qualitative results
extracted from Fig. \ref{fig1}. Similar trends are observed
for lower haloes masses (i.e. $5\times 10^{11} {\rm M}_\odot$) but not
shown in Fig. \ref{fig3} for sake of clarity.
More massive haloes, with masses $5\times 10^{12} {\rm M}_\odot$
and $> 10^{13} {\rm M}_\odot$, generally take more time to reach their final mass, 
so that our analysis of their mean density profiles can only start
from redshift $z=2.7$ and $z=2.3$ respectively (see Fig. \ref{fig1}). 
Due to the non-negligible past AGN activity they have already
experienced by that time, one can notice that H$_{\rm AGN}$ haloes of
these mass sub-samples already feature profiles 
that are quite flat and can even have mean slopes lower than their H$_{\rm DM}$ counterparts.
On the contrary,  at low redshift $\rho_{\rm AGN}$ profiles steepen
considerably, with $\gamma'$ reaching values well below -1, and close
to -1.5 by $z=0$, thereby confirming the visual trends highlighted in Fig. \ref{fig1}. 
It is also worth mentioning that at any given redshift, more massive haloes tend to have 
flatter inner density profiles than low mass ones, which confirms that
AGN feedback impact increases with halo mass.
To give explicit numbers, 
at $z=0$, density profiles of haloes with masses  $> 10^{13} {\rm M}_\odot$,  $5\times 10^{12} {\rm M}_\odot$, $10^{12} {\rm M}_\odot$
and $5\times 10^{11} {\rm M}_\odot$ boast an inner slope value of $\sim -1.3$, $\sim -1.5$, $\sim -1.8$ 
and $-2.1$ respectively (the latter is not shown in Fig. \ref{fig3}). 
Note that the range of [1-5] kpc is close to the simulation grid size (i.e. 1 kpc) where
 DM density profiles might not fully converge. This does not mean that the trends observed in Fig. \ref{fig3} are
necessarily wrong but one also need to consider slightly higher values in order to get a clear diagnostic.
Following Power et al. (2003),  a lower limit value of 5 kpc is recommended for the studied halo mass range,
though their analysis concerns pure dark matter simulations. Thus, in the following, we
also study the evolution of DM/stellar density profiles in the range of  [5-10] kpc for indicative purposes only.
  To this effect, we also show 
 in Fig. \ref{fig3} the evolutions of  $\gamma'_{\rm dm}$ but measured this time at [5-10] kpc and
 one can notice that similar evolutions and trends are obtained, 
though the variations from  H$_{\rm AGN}$ haloes are less pronounced, and therefore same 
overall conclusions can be drawn.

Fig. \ref{fig4}  shows the evolution
 of $\gamma'_{\rm dm}$ estimated either at [1-5] kpc or [5-10] kpc for the progenitors
 of dark matter haloes that have a  mass of $5\times 10^{12} {\rm M}_\odot$ either at $z=1.0$ or $z=0.0$.
Focusing first on the highest redshifts ($z >2$), H$_{\rm AGN}$ and H$_{\rm noAGN}$ profiles are clearly much steeper than H$_{\rm DM}$ ones.
As previously mentioned, this is mainly due to the
galaxy formation process: radiative cooling and 
subsequent star formation lead to a steepening of the DM density
profile of DM haloes due to adiabatic contraction
(Blumenthal et al. 1986; Gnedin et al. 2004),
in agreement with previous works (e.g., Gustafsson et al. 2006; Romano-D{\'{\i}}az et al. 2008
Abadi et al. 2010; Pedrosa et al. 2010; Tissera et al. 2010; Duffy et al. 2010).
Note that the mean slopes of H$_{\rm AGN}$ and H$_{\rm noAGN}$ halo
density profiles are very close
at $z\sim 5$ suggesting that AGN feedback has not had an important
impact yet by this redshift.
Fast forwarding to $z\sim 1.6$, whilst the  mean profile
of  H$_{\rm noAGN}$ haloes remains very steep with a slope close to -2, 
that of  their H$_{\rm AGN}$ counterparts is progressively flattened  by  AGN feedback, 
in good agreement with the results reported in Peirani et al. (2008) and  Martizzi et al. (2013).
Finally, from $z\sim 1.6$ to present times, $\rho_{\rm AGN}$ becomes slightly steeper
as the mean inner slope of H$_{\rm AGN}$ haloes progressively decreases by 0.1.
The two panels of Fig. \ref{fig4} explain the origin of the difference in
the slope of H$_{\rm AGN}$ halo density profiles at fixed halo mass for intermediate and low redshifts 
displayed in Fig. \ref{fig3}: H$_{\rm AGN}$ haloes tend to have steeper inner profiles
at low redshift because they have undergone a phase of ``cusp regeneration''.
For completeness sake, we recall that  additional refinement levels are triggered  at $z\sim 5$, $z\sim 1.5$
 and $z\sim 0.26$. They induce a sudden increase in star formation
 and black hole accretion at these different epochs as the gas is
 allowed to collapse to higher densities. These abrupt changes slightly affect the evolution of $\gamma'_{\rm dm}$ especially 
in the case H$_{\rm noAGN}$ haloes, as shown on Fig. \ref{fig4}.
Finally, it is worth mentioning that no significant difference is seen in the evolutions
of $\gamma'_{\rm dm}$ when using the ranges of [1-5] kpc or [5-10] kpc.

\begin{figure}
\rotatebox{0}{\includegraphics[width=\columnwidth]{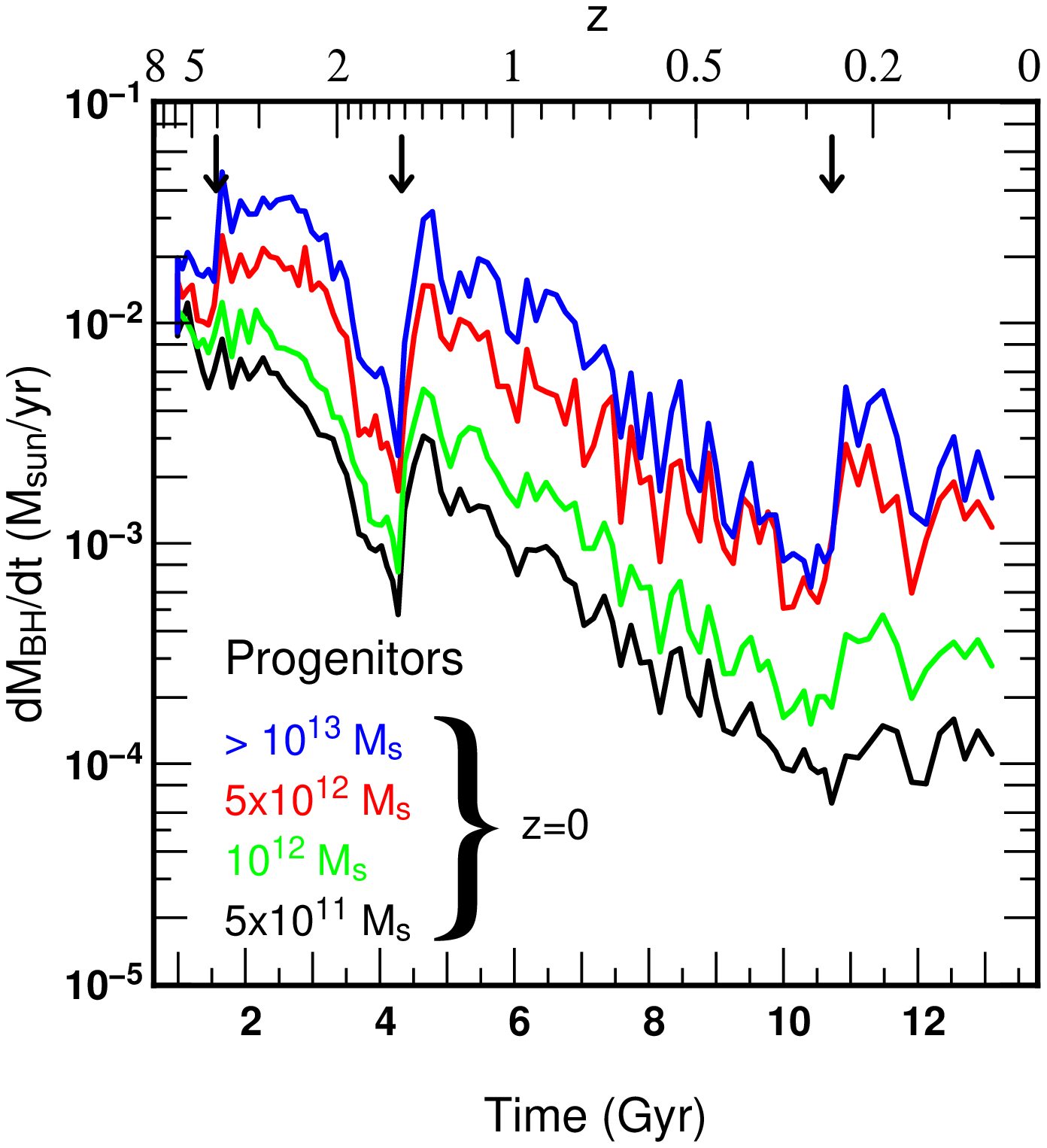}}
\rotatebox{0}{\includegraphics[width=\columnwidth]{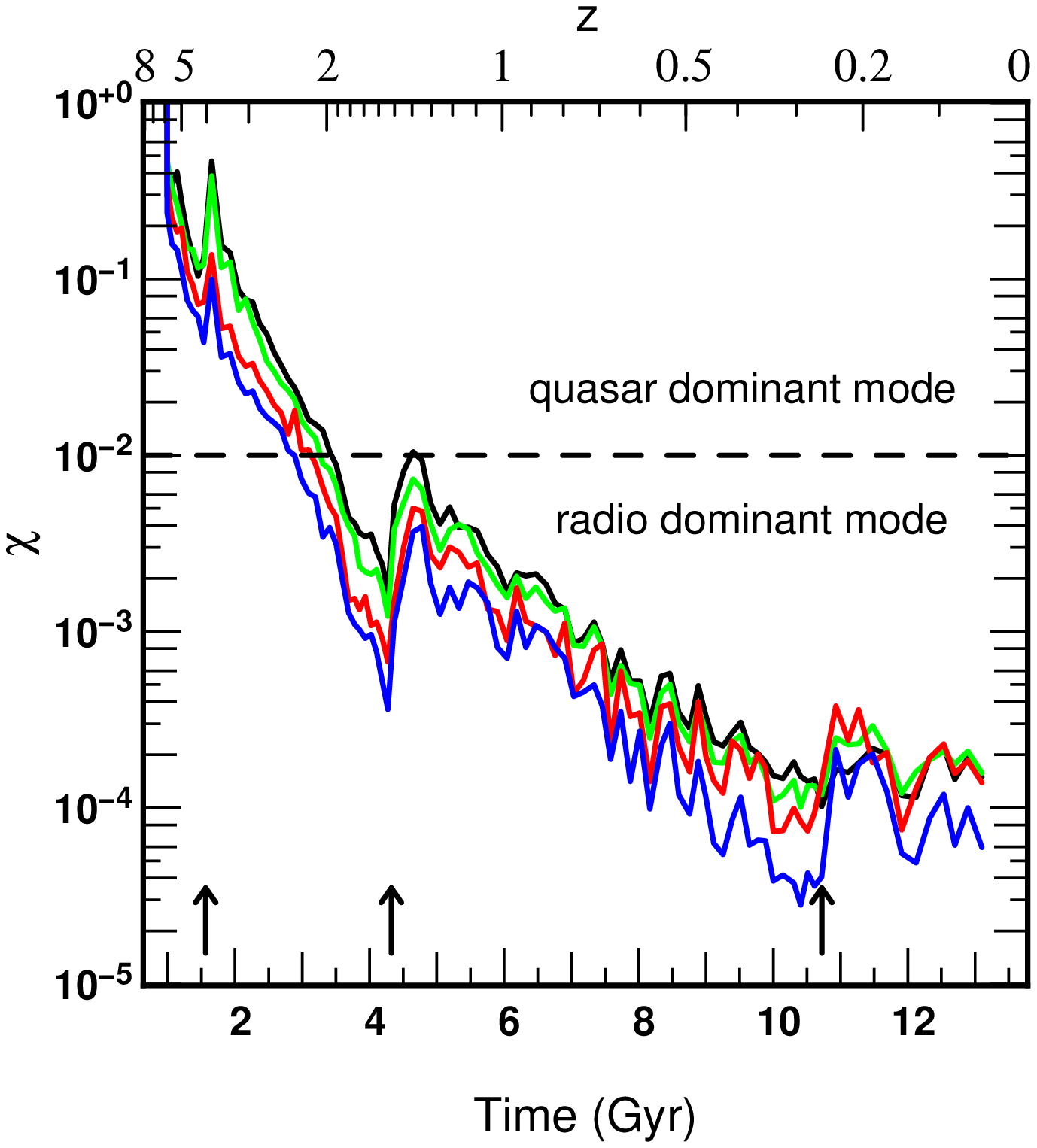}}
\caption{Upper panel: evolution of the median mass accretion onto black holes $\dot{M}_{\rm BH} \equiv dM_{\rm BH}/dt$
 for progenitors  of DM haloes of mass $5\times 10^{11} {\rm M}_\odot$ (black line), $10^{12} {\rm M}_\odot$ (green line),
 $5\times 10^{12} {\rm M}_\odot$ (red line) and $\geq 10^{13} {\rm M}_\odot$ (blue line) at $z=0$. Lower panel:
median Eddington ratio $\chi \equiv \dot{M}_{\rm BH}/\dot{M}_{\rm Edd}$ where  $\dot{M}_{\rm Edd}$ is
the Eddington accretion rate. Arrows indicate when an additional level
of refinement is added in the simulation. 
For a given halo mass, the  mass accretion onto black holes and therefore the AGN activity
 is much lower at low redshifts where the radio mode dominates.}
\label{fig5}
 \end{figure}

\begin{figure}
\rotatebox{0}{\includegraphics[width=8cm]{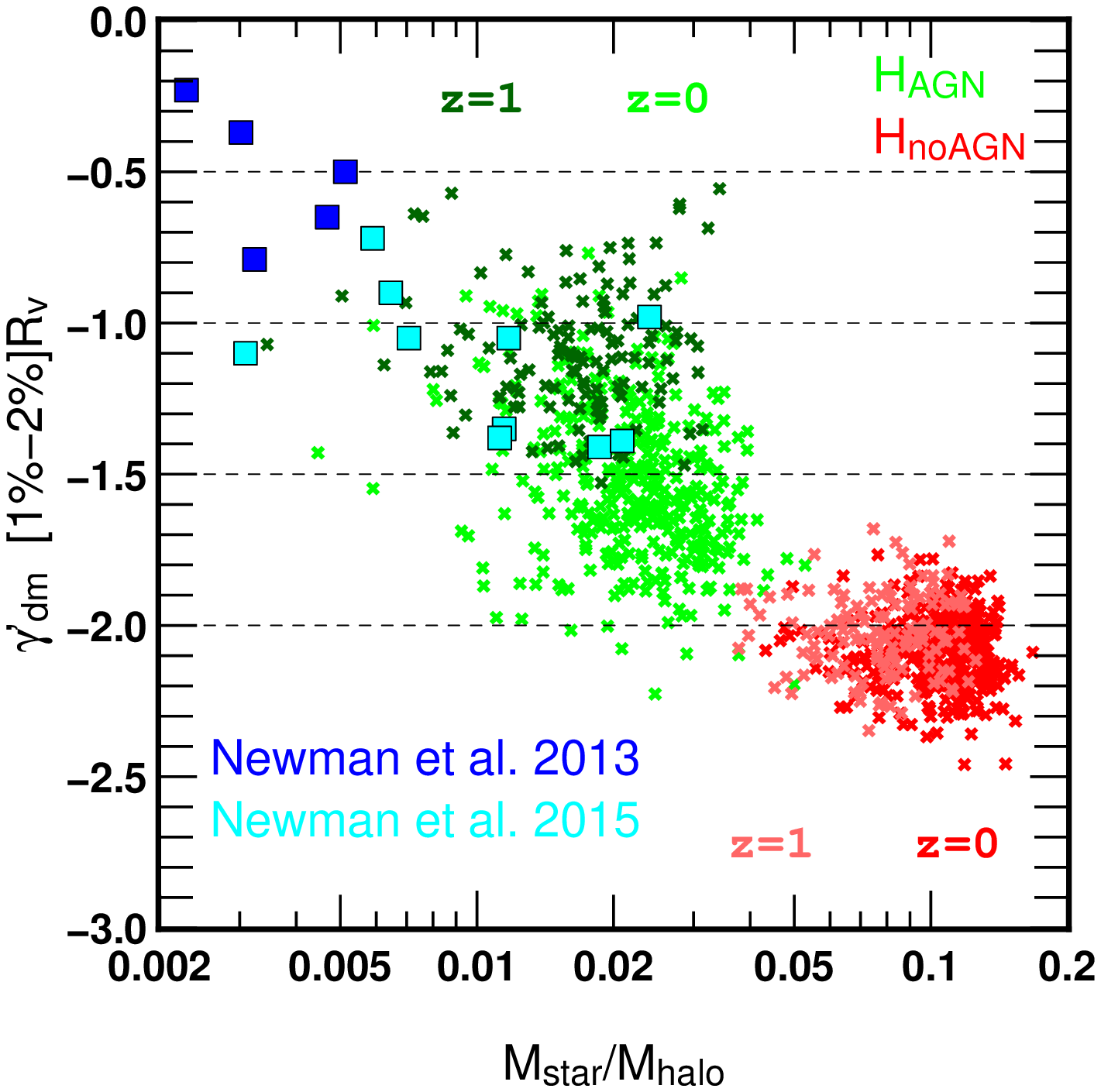}}
\rotatebox{0}{\includegraphics[width=8cm]{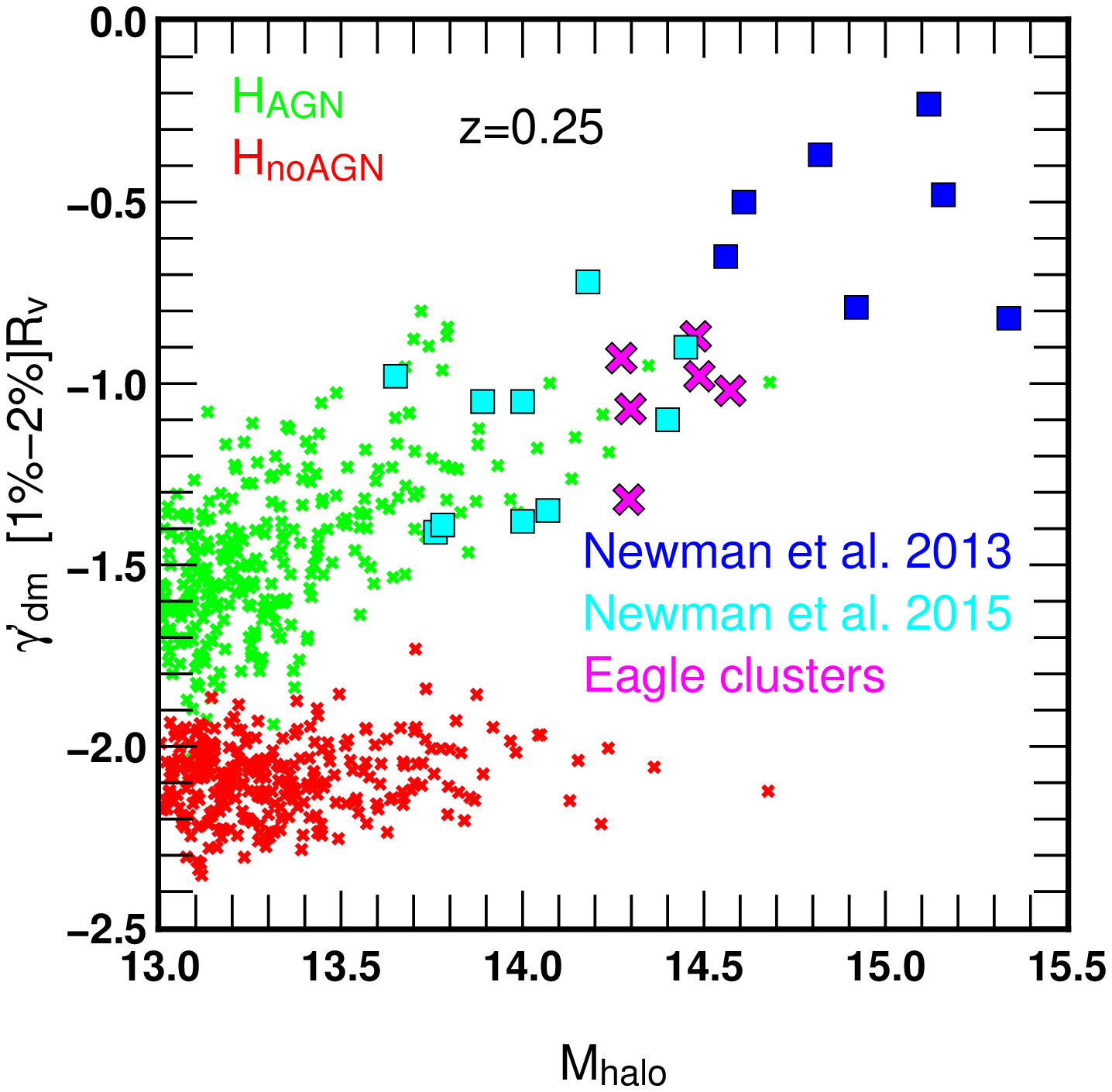}}
\caption{Upper panel: Variation of the mass-weighted density slope $\gamma'_{\rm dm}$ 
of DM profiles within [1-2]\% of the virial radius as a function of the
ratio $M_{\rm star}/M_{\rm halo}$ and halos mass for H$_{\rm AGN}$ haloes of
mass $\geq 10^{13} {\rm M}_\odot$ (green crosses) and matching  H$_{\rm noAGN}$ haloes (red crosses) identified at 
 $z=1$ (light colors) and $z=0$ (dark colors). Dark matter haloes with a higher $M_{\rm star}/M_{\rm halo}$ ratio
tend to have steeper inner profiles. Lower panel: Variation of $\gamma'_{\rm dm}$ as a function of the halo mass
for the same objects. In each panel, we also show the slope of the internal structure of DM haloes
for groups-scale lenses (cyan squares; Newman et al. 2015) and cluster-scale lenses (blue squares; Newman et al. 2013). Finally, we also plot results from Eagle clusters (Schaller et al. 2015b) in the lower panel. Here
most massive haloes tends to have a flatter profiles and theoretical predictions seems
to agree remarkably well with observations.}
\label{fig6}
 \end{figure}

Hence the evolution of the inner part of $\rho_{\rm AGN}$ does exhibit
three successive phases which are quantitatively differentiable.
A first condensation phase, from very high redshifts to $z\sim 5$
where dissipative galaxy formation processes and consequent adiabatic
contraction of dark matter haloes dominate. This stage is then followed by
a flattening phase, driven by important AGN activity until $z\sim 1.6$. Finally,
a second condensation phase or cusp regeneration occurs from $z\sim
1.6$ down to $z=0$ as AGN activity slowly subsides.
To understand this latter phase, one can study the evolution 
of AGN activity in the relevant dark matter halo sub-samples.
Indeed, an analysis of the evolution of gas mass accretion onto black holes (BHs)
 $\dot{M}_{\rm BH} \equiv dM_{\rm BH}/dt$ for progenitors of DM haloes
 taken from each mass sub-sample and displayed in Fig. \ref{fig5} reveals that for any given halo mass sample, 
$\dot{M}_{\rm BH}$ and therefore AGN activity progressively decrease with  time.
In spite of the noticeable bumps induced by the addition of new refinement
levels which artificially boost accretion,
$\dot{M}_{\rm BH}$ is one order of magnitude lower at $z=0$ than at $z\sim 1.4$.
The cusp regeneration phase measured in the evolution of  $\gamma'_{\rm dm}$ is 
 most probably related to the fading of AGN activity at lower
 redshifts. As emphasised by Peirani et al. (2008) and Martizzi et al. (2013),
the flattening of dark matter halo inner density profiles is due to a repetitive cycle of
rapid gas expansions driven by AGN feedback and slower contractions as
gas cools and falls back, which increases the velocity dispersion of the 
dark matter particles. If AGN activity progressively dwindles, 
this mechanism becomes less efficient at counterbalancing DM adiabatic contraction 
and thus keeping the DM density profiles flat. 
Fig. \ref{fig5} also shows the evolution of
 the Eddington ratio $\chi \equiv \dot{M}_{\rm BH}/\dot{M}_{\rm Edd}$ (where  $\dot{M}_{\rm Edd}$
 is the Eddington accretion rate)  for the same haloes.
These results suggest that the radio mode tends to be the dominant mode below $z\lesssim 2$.
A similar behaviour is observed when considering haloes of fixed mass, rather than progenitors. 
Such a redshift dependence of the dominant mode is also in agreement with Volonteri et al. (2016) who have studied in detail
 the cosmic evolution of black holes in the \hagnn simulation.
These authors found that the evolution of the luminosity function (LF)
in \hagnn is consistent with the existing observational determination of the bolometric LF
(Hopkins et al. 2007; Shankar et al. 2009; Ueda et al. 2014) provided there exists a transition at  $z\sim 2$
between quasar and radio dominated mode (see their Fig. 13).
Recall that in the \hagnn simulation, the explosive quasar mode consists in an  isotropic injection 
of thermal energy into the surrounding gas
while at low accretion rates, the more quiescent radio mode deposits AGN feedback energy into a bipolar outflow
(see Dubois et al. 2010 for a complete description).
It is therefore not very surprising that such a transition also
affects the evolution of inner dark matter halo profiles, 
since the radio mode of AGN feedback mainly prevents hot gas from cooling
and  eject little gas from the galaxy (Beckmann et al. 2017).

The cusp regeneration phase was already observed by Di Cintio et al. (2014) and
Tollet al. (2016) in hydrodynamical simulations
focusing on lower mass haloes (i.e. $M_{\rm halo}< 10^{12} {\rm M}_\odot$).
 These authors claimed that the density slope of DM haloes mainly 
depends on the ratio between stellar mass and total halo mass $M_{\rm
  star}/M_{\rm halo}$, with large ratios ($M_{\rm star}/M_{\rm
  halo}>0.01$) corresponding to contracted profiles.
Fig. \ref{fig6} shows the variation of the slope of DM halo density profiles, estimated this time
 within 1-2 \% of the virial radius, as a function of
 the ratio $M_{\rm star}/M_{\rm halo}$ for H$_{\rm AGN}$ haloes of mass $\geq 10^{13} {\rm M}_\odot$
 and matching  H$_{\rm noAGN}$ haloes. We find that in general, our haloes have values of $M_{\rm star}/M_{\rm halo}$
greater than $0.01$ and slopes steeper than -1, indicative of contraction. At $z=0$, H$_{\rm AGN}$ haloes have steeper density profiles and higher
 $M_{\rm star}/M_{\rm halo}$ values compared to haloes of the same mass at $z=1$. Moreover,  
 in the absence of AGN feedback, higher values of $M_{\rm star}/M_{\rm halo}\gtrapprox 0.05$ are 
obtained which are associated to steeper profiles ($\gamma'_{\rm dm}\sim -2$). 
 This suggests that $\gamma'_{\rm dm}$ is strongly correlated with
 $M_{\rm star}/M_{\rm halo}$, thus corroborating the previous findings
of Di Cintio et al. (2014) and Tollet al. (2016). However, it is worth mentioning that
their simulations did not include AGN feedback and therefore are
difficult to extrapolate to high mass haloes. Nonetheless, we remark that
our results  from H$_{\rm noAGN}$ haloes are in fair agreement with the predictions
of their fitting function for $M_{\rm star}/M_{\rm halo} > 0.05$.

Although a systematic comparison with observational data will be presented in 
a companion paper (Peirani et al. in prep), it is still very instructive to make some comparison
at this stage in order to understand whether the modelled processes
 are in the right efficiency ballpark or not. In Fig. \ref{fig6}, we also show the variation
of slope of the internal structure of DM haloes $\gamma'_{\rm dm}$ derived
 from several groups-scale lenses with $M_{200}\sim 10^{14} M_\odot$ at $<z>\sim 0.36$
(Newman et al. 2015) and cluster-scale lenses with  $M_{200}\sim 10^{15} M_\odot$ 
at $z\sim[0.2-0.3]$ (Newman et al. 2013). Although our samples are dominated by dark matter haloes
with a mass mass lower than $\sim 10^{14} M_\odot$, the theoretical trends obtained from \hagnn 
are quite consistent with observational expectations namely  dark matter haloes
 with a higher $M_{\rm star}/M_{\rm halo}$ ratio tend to have steeper inner profiles.
Such results are confirmed when studying the variation of $\gamma'_{\rm dm}$ as
a function of halo mass displayed in the lower panel of  Fig. \ref{fig6}. 
In this case, more massive DM haloes tend to have flatter density profiles. 
These figures also suggest
that the inclusion
of AGN feedback leads to a much better agreement with observational values and trends which is
one of the main conclusion of our forthcoming study.
Note also that it is quite promising that our theoretical predictions seem to be in good agreement
with the Eagle simulation ones (Schaller et al. 2015b). In this regard, we only show results from
the Eagle clusters but a similar nice agreement is also obtained for lower mass range haloes
(see, for instance, Fig. 4 of Schaller et al. 2015).

\subsection{Evolution of the gap between $\rho_{\rm AGN}$, $\rho_{\rm noAGN}$ and $\rho_{\rm DM}$}

In this section, we study the evolution of $A_{\rm DM}$ and $A_{\rm noAGN}$ 
defined by equation \ref{equ2}
which monitors how the relative gaps between $\rho_{\rm AGN}$, $\rho_{\rm noAGN}$ and $\rho_{\rm DM}$ evolve.
The variations of  $A_{\rm DM}$ and $A_{\rm noAGN}$  are thus expected
to provide complementary information
to the evolution of the slope of dark matter halo density profiles
previously studied.

\begin{figure}
\rotatebox{0}{\includegraphics[width=\columnwidth]{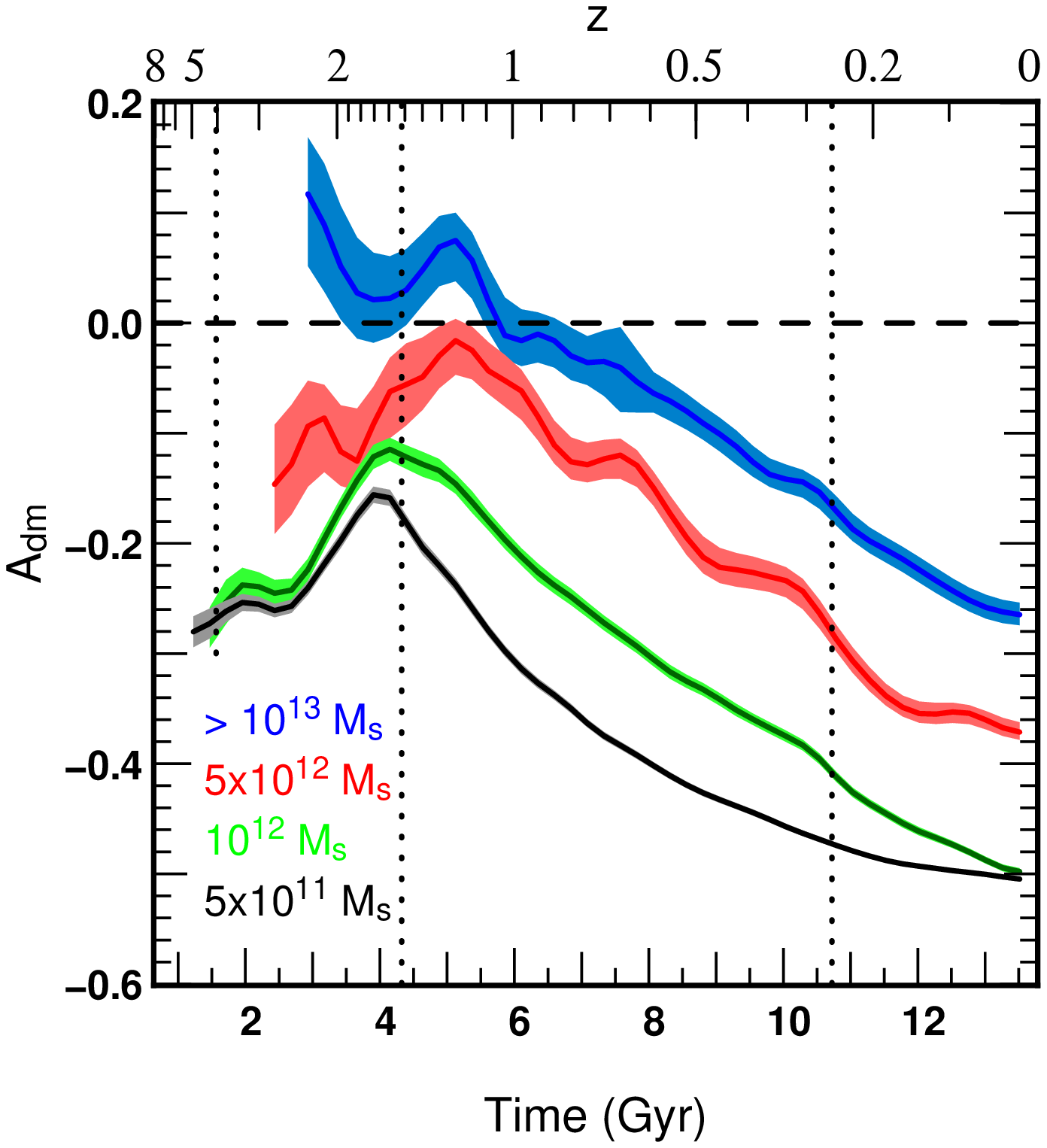}}
\rotatebox{0}{\includegraphics[width=\columnwidth]{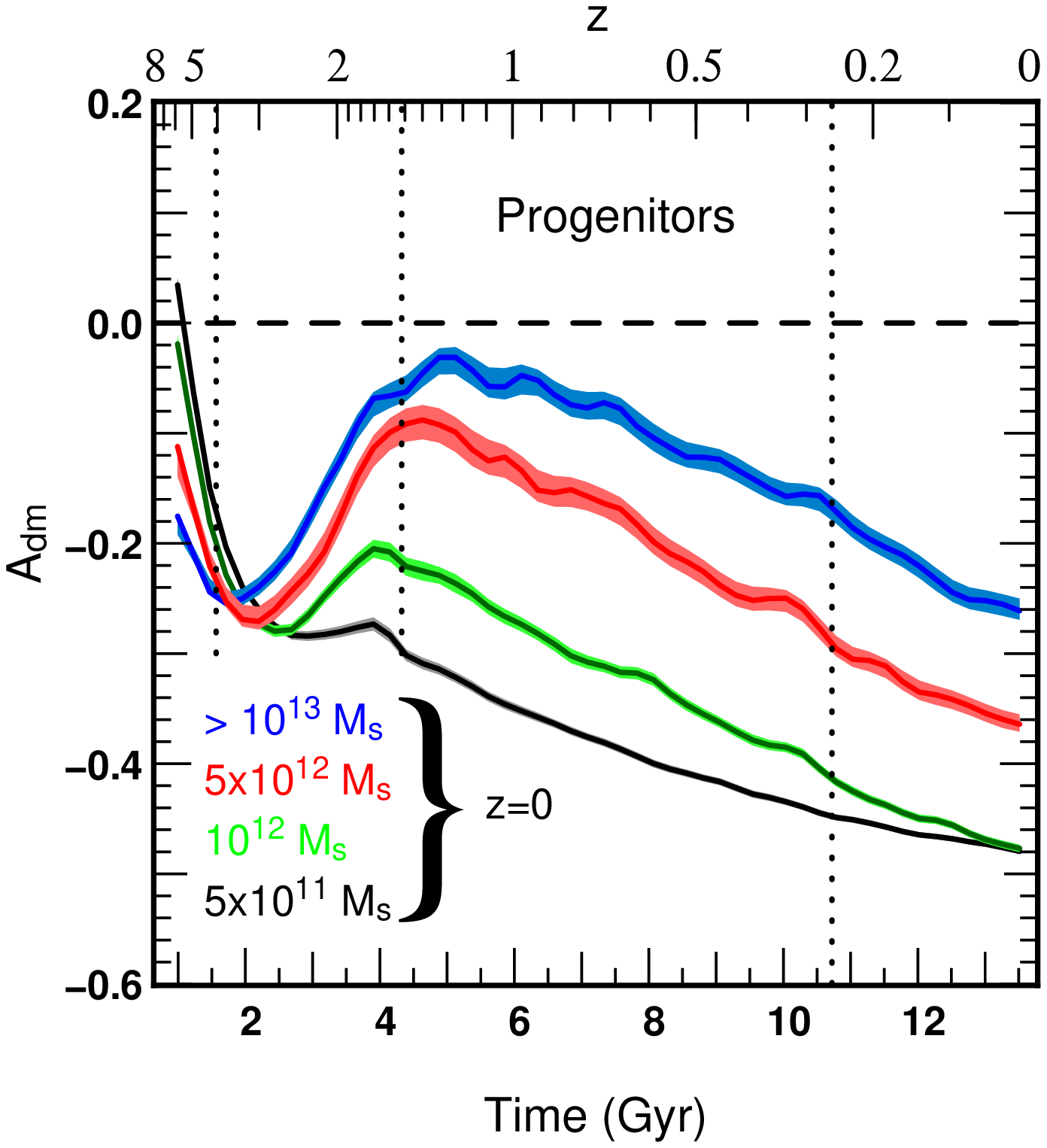}}
\caption{Time evolution of $A_{\rm DM}$ which measures the gap
between H$_{\rm AGN}$ and matched H$_{\rm DM}$ dark matter halo density profiles in the range of [1-10] kpc. 
The upper panel considers haloes within fixed mass intervals,
independently of cosmic time while
the lower one follows the evolution of $A_{\rm DM}$ for the
progenitors of haloes of mass $\sim 5\times10^{11}$, $\sim 10^{12}$,
 $\sim 5\times10^{12}$ and $\geq 10^{13} {\rm M}_\odot$ at $z=0$. The vertical dashed lines
indicate when additional levels of refinement are triggerred. The
shaded areas represent the error
on the mean.
Note that similar trends are obtained when estimating $A_{\rm DM}$ in the range of [1-5] kpc or
  [5-10] kpc}.
\label{fig7}
 \end{figure}

\subsubsection{H$_{\rm AGN}$ vs H$_{\rm DM}$}

Let us start with  the relative evolution of  $\rho_{\rm DM}$ and $\rho_{\rm AGN}$.
Fig. \ref{fig7} shows the variations of $A_{\rm DM}$ for the
haloes of our four considered mass intervals at $z=0$ (top panel) and 
their progenitors (bottom panel).
Three successive phases can clearly be identified again, particularly
when looking at the bottom panel of Fig. \ref{fig7}.
At very high redshift,  $A_{\rm DM}$ tends to 0 as 
galaxy formation has yet to significantly affect the inner structure
of dark matter haloes. 
Then, from very high redshift to $z\sim 3$, $A_{\rm DM}$ becomes
progressively more and more negative
due to rapid galaxy formation and subsequent dark matter adiabatic
contraction. From $z\sim 3$ to $z\sim [1.6-1.2]$,  $A_{\rm DM}$ is
increasing whilst retaining its negative value.
This suggests again that AGN activity is reducing the central density (and the inner density slope)
 of  H$_{\rm AGN}$ haloes which results in reducing the gap w.r.t. the matching H$_{\rm DM}$ haloes density profiles.
Finally, from  $z\sim [1.6-1.2]$ to $z=0$, $A_{\rm DM}$ decreases
again, always remaining negative. The gap between the inner
 density profiles of  H$_{\rm AGN}$ and H$_{\rm DM}$ haloes thus progressively becomes more and more important since this time
the central density (and the inner density slope) of H$_{\rm AGN}$ haloes is increasing (phase of cusp regeneration).
Note that the second phase (i.e. flattening of H$_{\rm AGN}$ DM halo density profiles) occurs sooner and lasts longer for
more massive objects. Furthermore, the third phase  occurs
later for more massive objects. This reinforces our conclusions based
on the evolution of inner profile density slopes that AGN feedback
impacts more durably and significantly the most massive haloes.

A similar conclusion can be drawn when studying the time evolution of
$A_{\rm DM}$ for haloes with a fixed mass, independent of cosmic time, as 
displayed in the top panel of Fig. \ref{fig7}.
$A_{\rm DM}$ is increasing at high
redshifts ($5>z>2$) for the first three lower mass intervals while keeping negative values.
These black, green and red curves reach a maximal value at redshifts around 1.6, 1.6 and and 1.2
respectively: AGN activity has efficiently reduced the central density
as well as flattened the profiles of H$_{\rm AGN}$ haloes, leading to the reduction of
the gap between inner density profiles of H$_{\rm AGN}$ and matching H$_{\rm DM}$ haloes.
Note that for these three mass intervals,  $A_{\rm DM}$ is already negative 
at high redshift due to DM adiabatic contraction induced by galaxy
formation having already taken place in such massive haloes.
 Then, after each peak, $A_{\rm DM}$ monotonically decreases until $z=0$ (phase of cusp regeneration).
For the most massive haloes ($\geq 10^{13} {\rm M}_\odot$), $A_{\rm DM}$
is mainly decreasing but remains positive until $z\sim 1$ indicating
that H$_{\rm AGN}$  haloes of this mass at higher redshifts have lower
central density (and also flatter density profiles) than their
H$_{\rm DM}$  counterparts. This also reinforces that 
AGN activity has a stronger impact on the inner DM density profiles of the most massive haloes.

\begin{figure}.
\begin{center}
\rotatebox{0}{\includegraphics[width=\columnwidth]{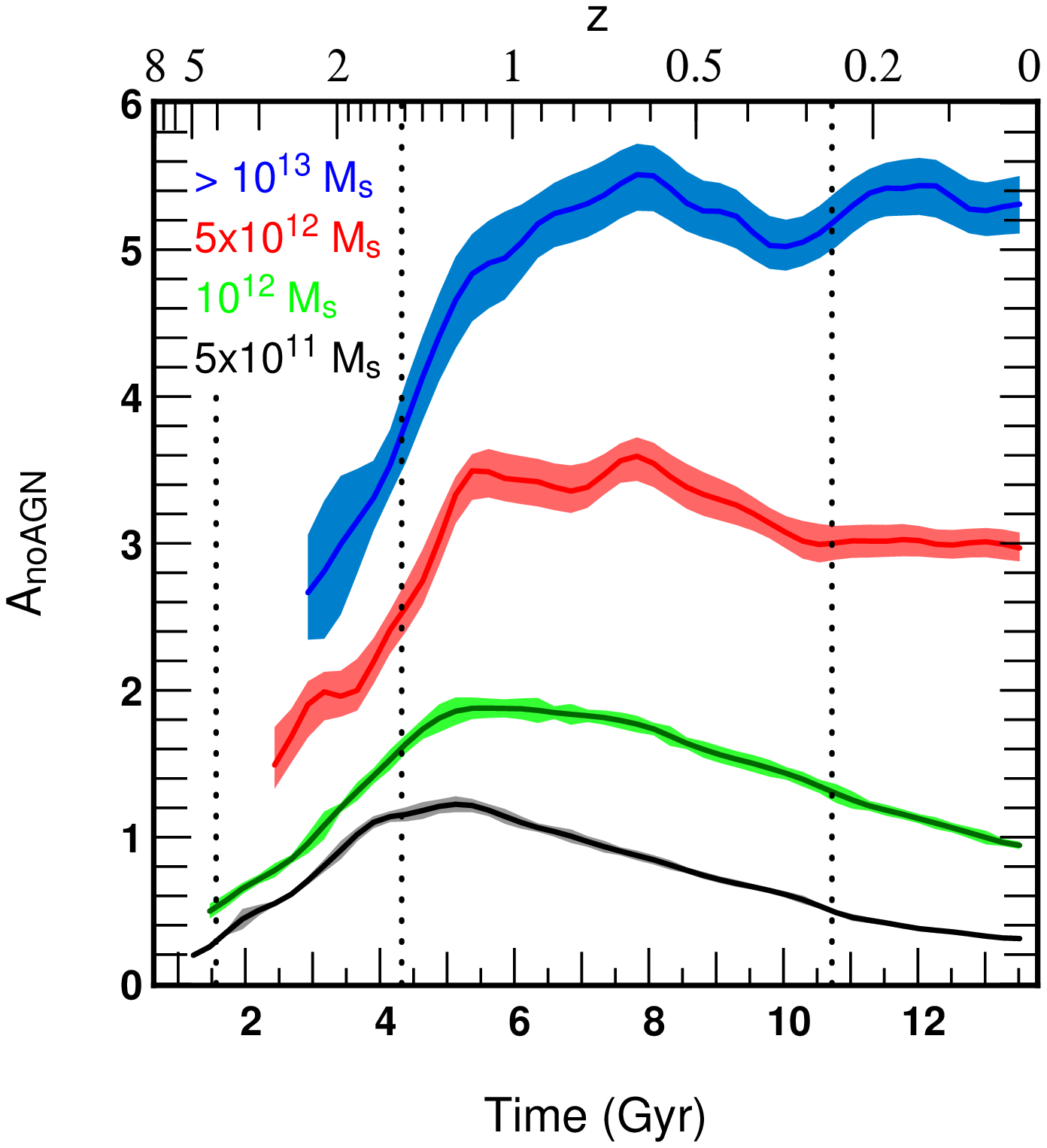}}
\rotatebox{0}{\includegraphics[width=\columnwidth]{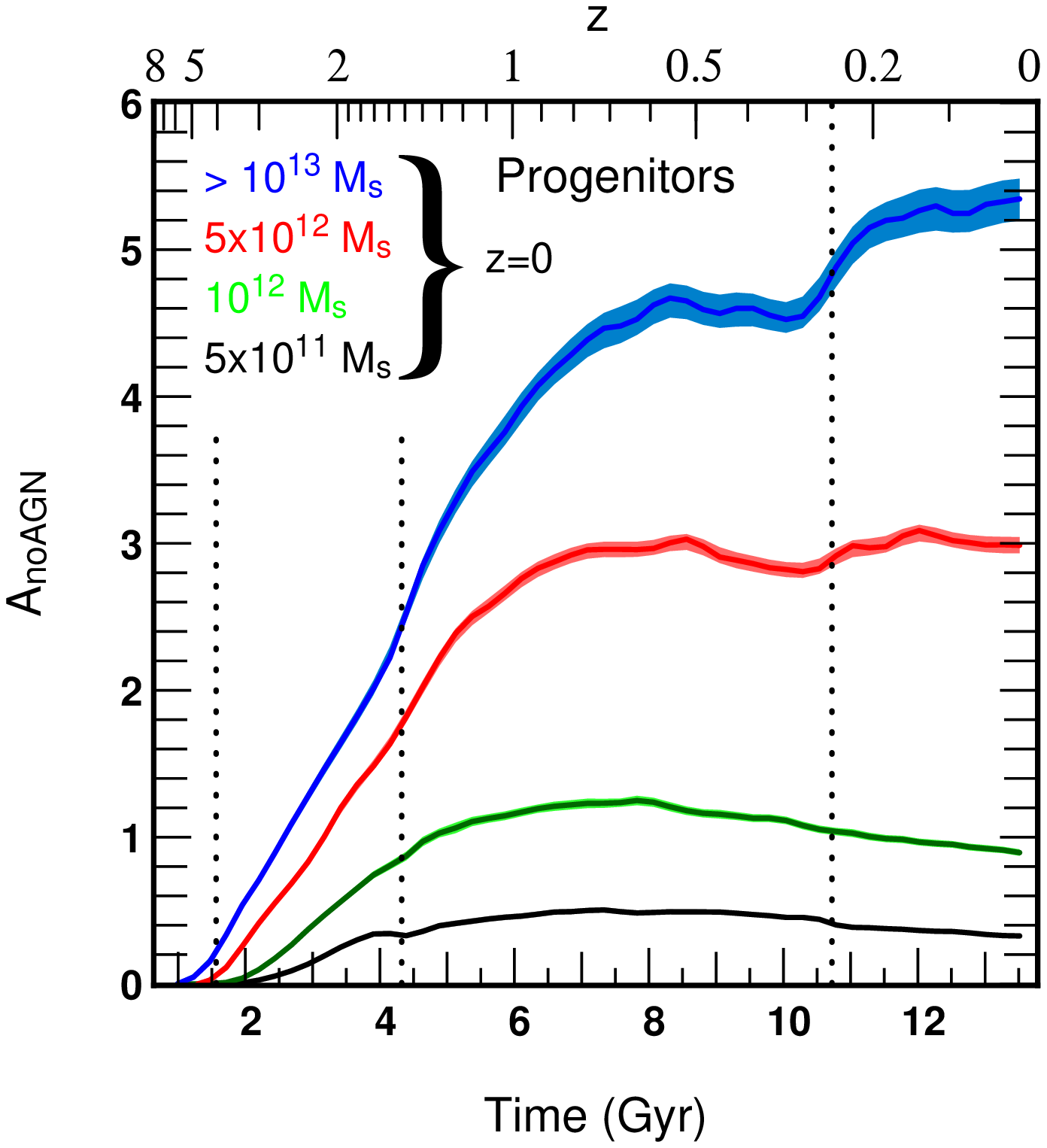}}
\caption{Time evolution of $A_{\rm noAGN}$, which measures the gap
between H$_{\rm AGN}$ and matching H$_{\rm noAGN}$ DM halo density profiles in the range of [1-10] kpc. 
Upper and lower panels consider haloes within fixed mass intervals across cosmic time 
and the progenitors of haloes with masses of $\sim 5\times10^{11}$, $\sim 10^{12}$,
 $\sim 5\times10^{12}$ and $\geq 10^{13} {\rm M}_\odot$ at $z=0$ respectively. The vertical dashed lines
indicate when additional levels of refinement are triggerred in the
simulation. Shaded areas represent the error on the mean.
Here again, similar trends are obtained when estimating $A_{\rm noAGN}$ in the range of [1-5] kpc or
  [5-10] kpc}.
\label{fig8}
\end{center}
 \end{figure}

\subsubsection{H$_{\rm AGN}$ vs H$_{\rm noAGN}$}

AGN feedback manifests itself through two effects. First, it reduces star formation rates 
in massive galaxies and therefore the amount of stellar material present in their central parts.
 Second, as we have previously discussed, it flattens the DM host halo density profile,  especially at intermediate
 redshifts. As a result, the central dark matter profiles of H$_{\rm AGN}$ haloes should always be less steep and have 
lower central values when compared with their H$_{\rm noAGN}$ halo
counterparts. In other words, $A_{\rm noAGN}$ must always be positive.

Fig. \ref{fig8} shows the variation of $A_{\rm noAGN}$ for both progenitors and haloes within fixed 
mass intervals and confirms again the trends previously pointed out.
In the top panel, for the three less massive halo mass intervals,  the 
gap between the DM halo density profiles in H$_{\rm AGN}$ and  H$_{\rm noAGN}$ 
 first increases until it reaches a maximum value for $z$ close to
 1.2. It then slightly decreases as the redshift progresses towards $z=0$. 
For the most massive halos, with masses $\geq 10^{13} {\rm M}_\odot$, the gap
is essentially always increasing until $z=0.8$ after which it remains
constant until $z=0$. This is  consistent with our previous result: the effect of AGN feedback (quasar mode)
is more efficient at high redshift, leading to a rapid increase of the gap between the  H$_{\rm AGN}$ and
  H$_{\rm noAGN}$ DM halo profiles. Then, in a longer and later
  phase  between $1.2 \geq z \geq 0$, AGN activity diminishes, 
and the gap narrows slightly.

The bottom panel of Fig. \ref{fig8}  presents the variations of $A_{\rm noAGN}$
focusing on the evolution of the DM halo profiles of the progenitors of haloes with a mass of
 $5\times10^{11} {\rm M}_\odot$, $10^{12} {\rm M}_\odot$,
 $5\times10^{12} {\rm M}_\odot$ and $\geq 10^{13} {\rm M}_\odot$ at $z=0$. Here again, at very high redshift,
$A_{\rm noAGN}$ is close to 0 for all considered mass intervals since AGN feedback 
has not yet kicked in. The gap between H$_{\rm AGN}$ and  H$_{\rm noAGN}$ haloes then increases until $z\sim [1-0.6]$ and stays roughly constant afterwards. 
It is worth mentioning again that adding an extra refinement level at 
$z\sim 1.5$ and $z\sim 0.25$ causes spurious artefacts to appear in
the otherwise rather smooth evolution of $A_{\rm noAGN}$. This corresponds to a sudden better force resolution and enhanced
 gas condensation which has a more dramatic effect in {\sc \small
   Horizon-noAGN} than in {\sc \small Horizon-AGN}, 
where it is somewhat compensated by a simultaneous rise in AGN activity.

\begin{figure}
\rotatebox{0}{\includegraphics[width=\columnwidth]{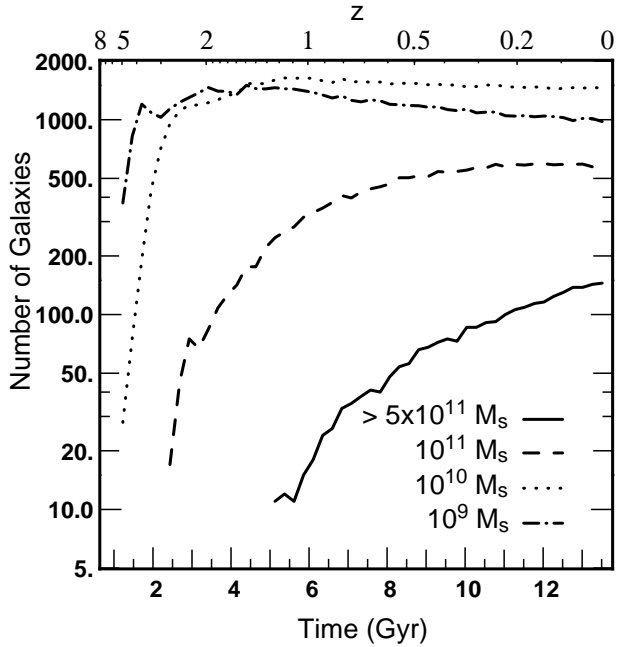}}
\caption{The number of galaxies matched between the two hydrodynamical simulations for
our four distinct fixed mass intervals namely $\sim 10^{9}  {\rm M}_\odot$ (dotted-dashed line), $\sim 10^{10} {\rm M}_\odot$ (dotted line),
 $\sim 10^{11}  {\rm M}_\odot$ (dashed line) and
$\geq 5\times10^{11} {\rm M}_\odot$ (solid line). We derived statistics only when 10 objects can be
identified at a specific redshift.}
\label{fig10bis}
 \end{figure}

\begin{figure*}
\rotatebox{0}{\includegraphics[width=16.5cm]{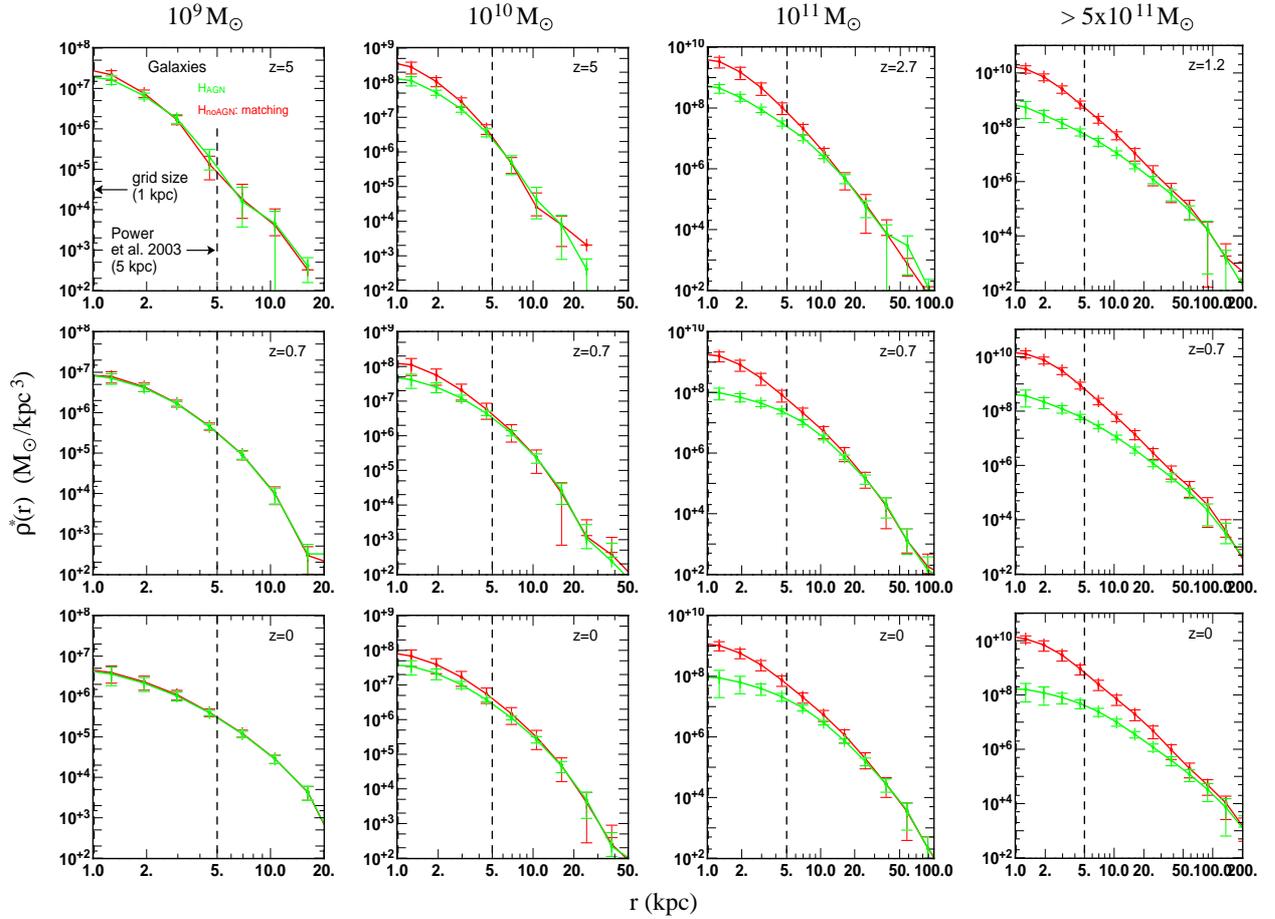}}
\caption{Evolution of the mean stellar density profiles of galaxies
 extracted  from H$_{\rm AGN}$ (green lines) and H$_{\rm noAGN}$
 (matching, red lines). We divide our galaxy sample on four fixed
mass intervals: $\sim 10^{9}  {\rm M}_\odot$ (first column), $\sim 10^{10} {\rm M}_\odot$ (second column),
 $\sim 10^{11}  {\rm M}_\odot$ (third column) and
$\geq 5\times10^{11} {\rm M}_\odot$ (fourth column). As for the DM
haloes, three different epochs are  considered: high redshift (first row),
$z=0.7$ (second row) and $z=0$ (third row). The two vertical dashed lines
at $r=1$ kpc and $r=5$ kpc are also displayed to indicate respectively the simulation grid size and a 
recommended resolution limit following Power et al. (2003). The error bars correspond to the dispersion. 
The discrepancy between density profiles is all the more
important than galaxies are massive. Moreover, the mean H$_{\rm AGN}$
stellar density profiles remain quite flat at low redshift.}
\label{fig9}
 \end{figure*}


\section{Stellar density profiles}
\label{section_galaxies}

\subsection{General trends}

In the following, we compare the stellar density profiles of galaxies in the same way as  
 we previously did for the dark matter component:  
 we  use a systematic  
 (AGN) galaxy  to (noAGN) galaxy object-to-object comparison using our matching procedure.
We also focus our analysis on  four (H$_{\rm AGN}$) galaxy stellar mass intervals: 
$10^{9} (\pm 10\%)  {\rm M}_\odot$,
$10^{10}  (\pm 10\%)  {\rm M}_\odot$,
$10^{11}  (\pm 10\%)  {\rm M}_\odot$ and
$\geq 5\times10^{11} {\rm M}_\odot $.
Fig. \ref{fig10bis} indicates the number of galaxies identified in each mass bin and for 
each considered redshift.

Fig. \ref{fig9} displays spherically averaged galaxy stellar mass
density profiles, $\rho_{\rm *,AGN}$ and matching $\rho_{\rm *,noAGN}$,
 derived for each of the four stellar mass bins and 
at three different redshifts. We again indicate in each panel, the simulation grid size (1 kpc)
and a lower resolution limit (5 kpc) recommended by Power et al. (2003).  
For the lowest mass interval
 (i.e. $10^{9}  {\rm M}_\odot$), we notice no significant
difference between H$_{\rm AGN}$ and H$_{\rm noAGN}$ galaxy density
profiles: this is not very surprising as AGN feedback is not thought to be effective in such low mass objects. 
For the three most massive mass intervals, clear gaps 
between  $\rho_{\rm *,AGN}$ and  $\rho_{\rm *,noAGN}$  can be seen in
the central regions, and up to a radius of 50 kpc
 for the most massive galaxies. 
 Moreover, the gap between inner density profiles is all the more important 
 than the galaxies are massive, and increases slightly with time.
  Similarly to what we observe for the dark matter component, 
 stellar density profiles with and without AGN feedback are  similar at large radii.
However, in contrast to what happens for dark matter haloes,  H$_{\rm  AGN}$ galaxy density profiles remain quite flat at low redshift, as
already suggested by Fig. \ref{fig2}. Such a behaviour is also
more consistent with observations (e.g. Kormendy et al. 2009).

Note that, at a given mass, galaxies are in general more extended at low
redshifts because they experience more mergers that tend
to spread material at large radii, which increase their effective radius over time
(Khochfar \& Silk 2006; Bournaud, Jog \& Combes 2007; 
Naab, Johansson \& Ostriker 2009; Peirani et al 2010; Oser et al. 2010, 2012; 
Shankar et al. 2013; Welker et al. 2017; Rodriguez-Gomez et al. 2016).
This effect is more pronounced with AGN feedback since the in situ star formation
 is regulated by AGN activity, at the benefit of the accreted stellar mass in the overall
 stellar mass budget (Dubois et al. 2013, 2016).

\begin{figure}
\rotatebox{0}{\includegraphics[width=\columnwidth]{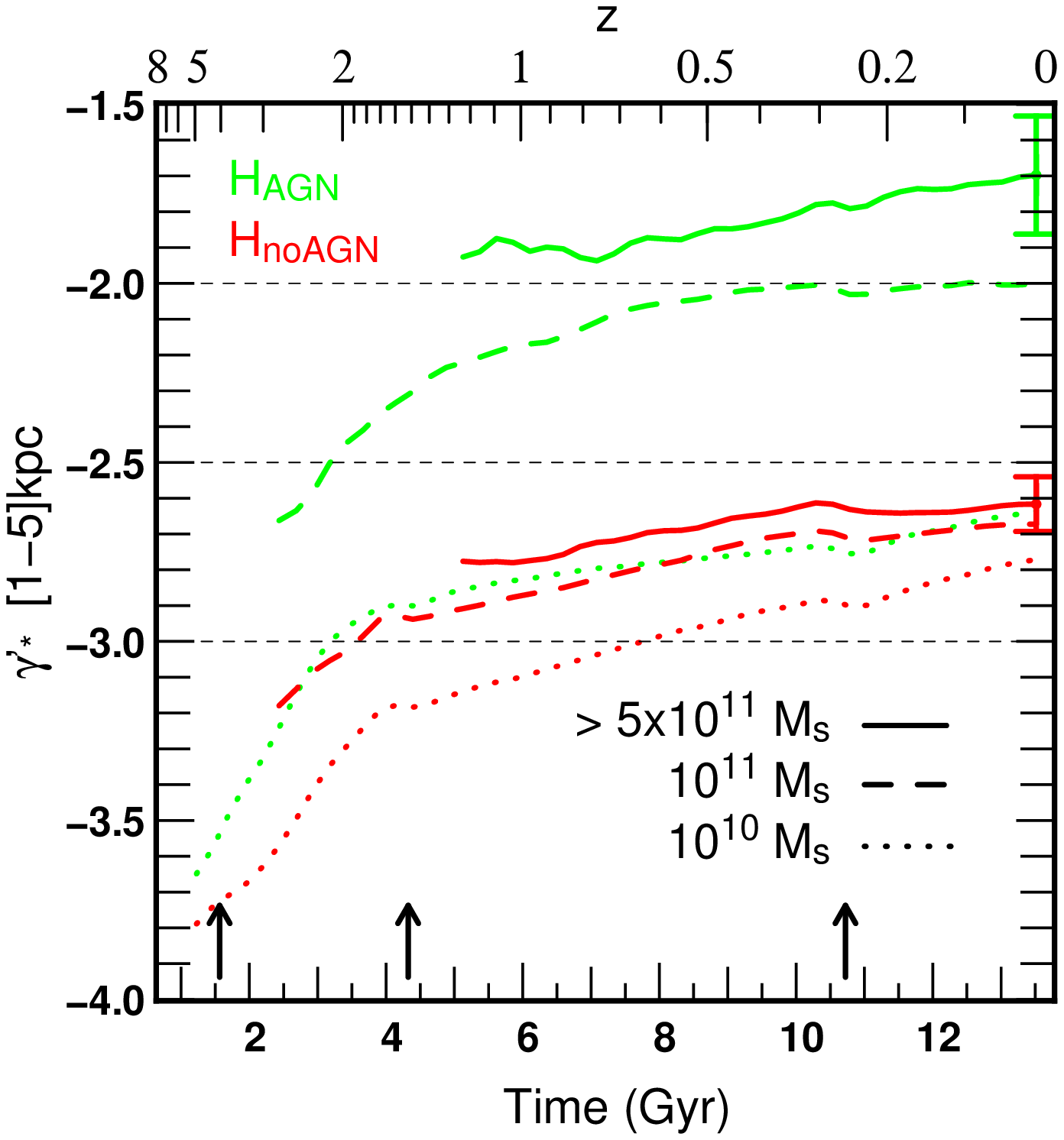}}
\rotatebox{0}{\includegraphics[width=\columnwidth]{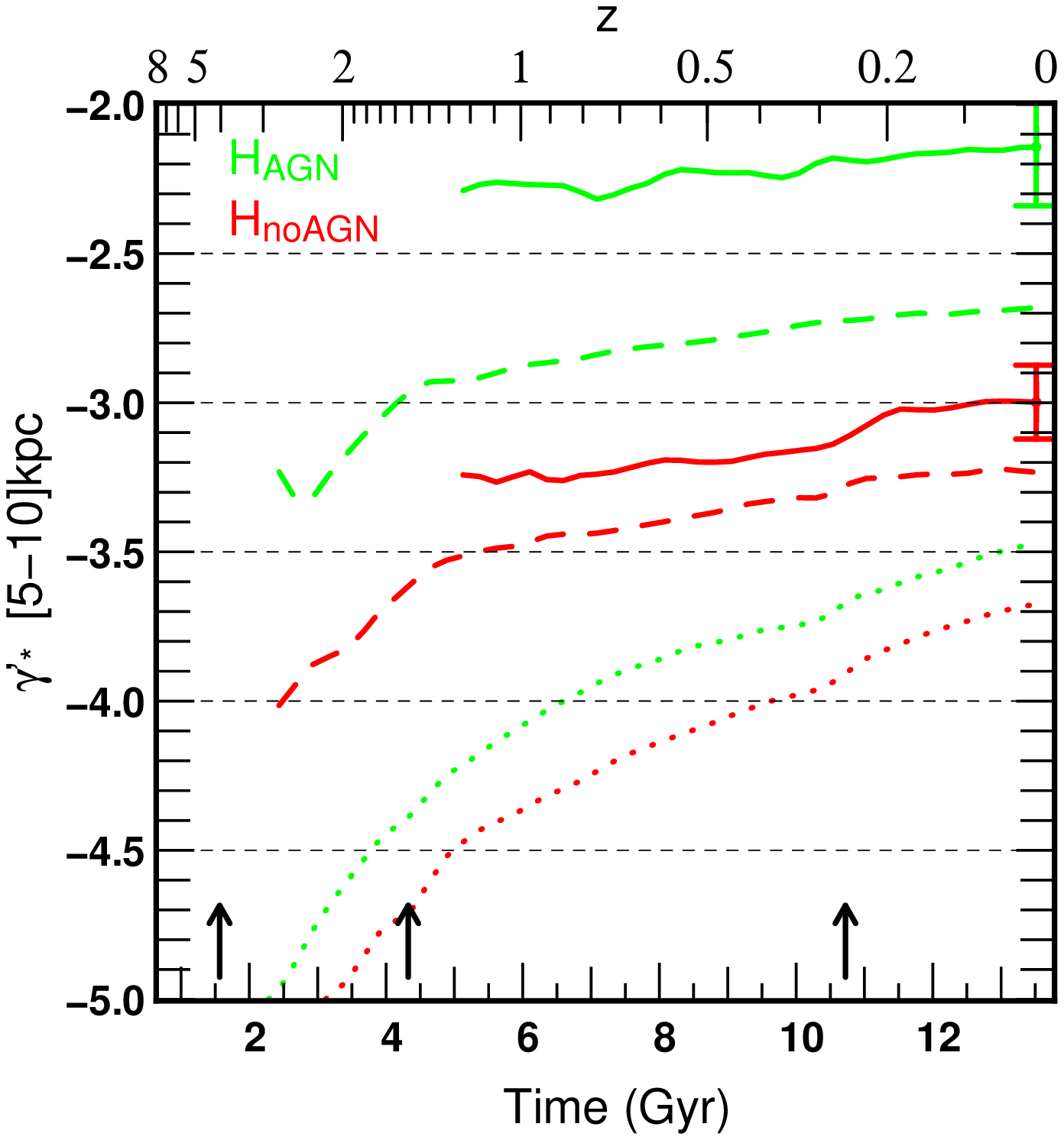}}
\caption{Evolution of the galaxy mass-weighted density slope $\gamma'_*$ estimated
within [1-5] kpc (upper panel) and [5-10] kpc (lower panel). 
We show results for the three more massive H$_{\rm AGN}$ galaxy mass
intervals (green colour) and matched  H$_{\rm noAGN}$ galaxies (red colour).
The three arrows indicate the times when an additional 
refinement level is added to the simulations. Typical standard
deviations are represented by vertical error bars at $z=0$. 
AGN feedback significantly and monotonically flattens the inner stellar density profiles of galaxies.}
\label{fig10}
 \end{figure}

In order to derive a more quantitative evolution, we first study the
evolution of the galaxy mass-weighted density slope $\gamma'_*$ measured
within [1-5] kpc and [5-10] kpc for H$_{\rm AGN}$ galaxies within a fixed mass bin,
independent of cosmic time,  and their matching H$_{\rm noAGN}$ counterparts.
 Those evolutions are displayed in Fig. \ref{fig10}.
For  galaxies with masses of $10^{10} {\rm M}_\odot$, $10^{11} {\rm M}_\odot$  and  $> 5\times 10^{11} {\rm M}_\odot$ H$_{\rm AGN}$,
 we see clear differences when AGN is included or not.
Stellar density profiles of H$_{\rm AGN}$ galaxies always display
shallower inner slopes than their  H$_{\rm noAGN}$ counterparts. 
Therefore, as was measured for the DM component, 
AGN feedback tends to flatten the inner stellar density profiles of massive galaxies.
We also note that H$_{\rm AGN}$ mean stellar density slopes increase
rapidly  at high redshift ($z > 1.5$) and then stall at lower $z$,
as a consequence of the evolution of
black holes growth and AGN activity history reported in Fig. \ref{fig5}.
Finally, the most massive  H$_{\rm AGN}$ galaxies
tend to have a  flatter inner profiles at any given redshift than
their lower mass equivalents, which suggest again that AGN feedback
plays a more important role in shaping more massive objects. 
Note that no particular difference are obtained when estimating
$\gamma'_*$ within [1-5] kpc or [5-10] kpc.

\begin{figure}
\rotatebox{0}{\includegraphics[width=\columnwidth]{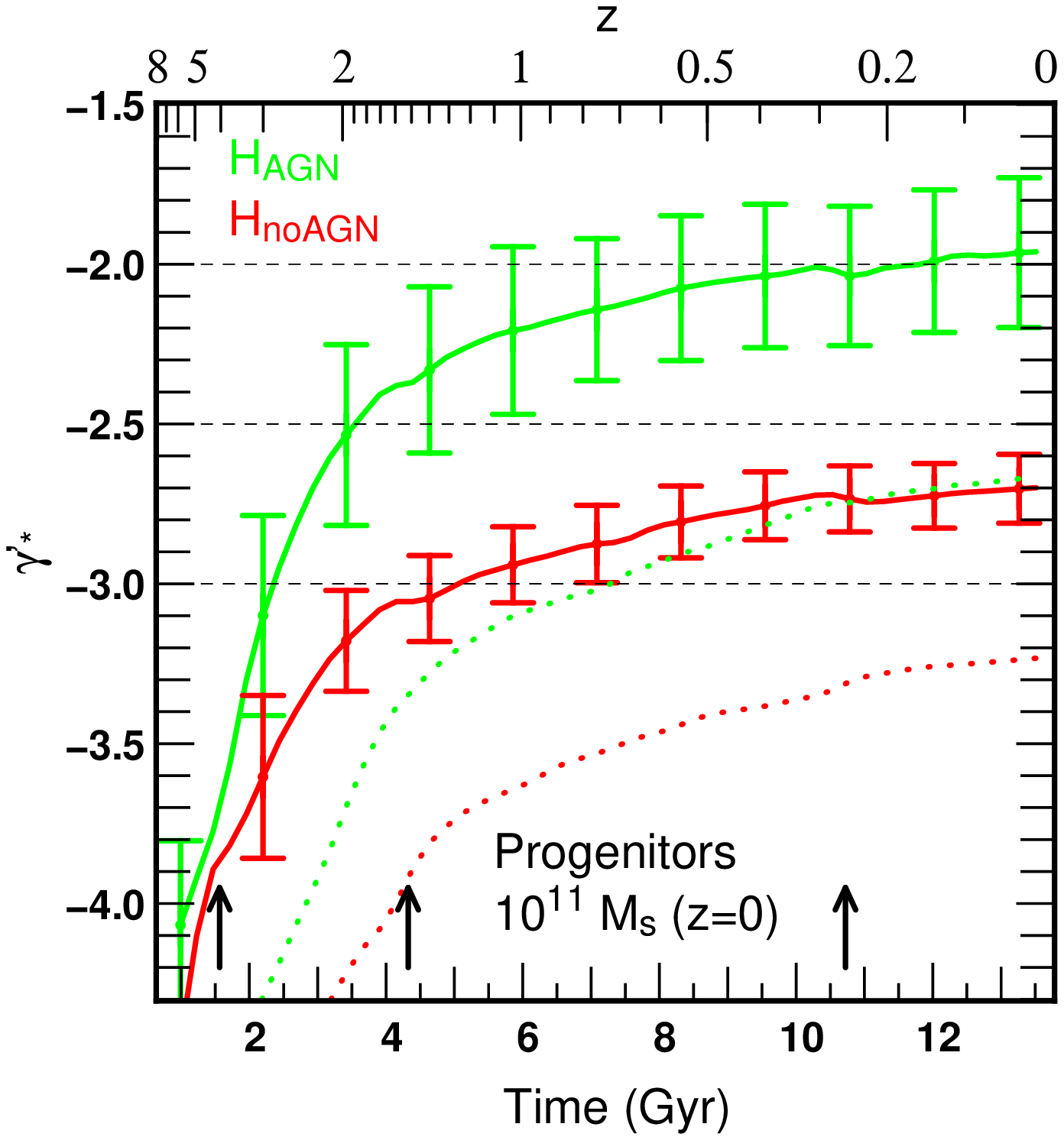}}
\caption{Time evolution of the mass-weighted density slope $\gamma'_*$ 
of the progenitors of galaxies with masses $10^{11} {\rm
  M}_\odot$ at $z=0$ estimating within [1-5] kpc (solid lines) or [5-10] kpc (dotted line).
Results from  H$_{\rm AGN}$ galaxies and their matching H$_{\rm noAGN}$ counterparts
are displayed in green and red colours respectively.
The three arrows represent epochs where an additional 
refinement level is added in the simulations. Error bars indicate the
standard deviations. H$_{\rm AGN}$ galaxy stellar density
profiles are rapidly flattened until $z\sim 1$. Without AGN feedback, galaxy density profiles 
always remain relatively steep.}
\label{fig11}
 \end{figure}

We  derived in Fig. \ref{fig11} 
the evolution of $\gamma'_*$ (estimated within either at [1-5] kpc or [5-10] kpc)
 for the progenitors of galaxies of mass
$10^{11} {\rm M}_\odot$ at $z=0$. 
At high redshift ($z>5$), 
since AGN activity has not yet picked up, the density slopes of H$_{\rm AGN}$ and H$_{\rm noAGN}$ 
galaxies  are very similar. Then, from $z\sim 5$ to $z\sim 1$, 
 AGN feedback strongly flattens the density profiles of H$_{\rm AGN}$ galaxies.
 At lower redshifts, the inner stellar density profile slopes remain
 almost constant or increase slightly
which confirms trends seen in Fig. \ref{fig2}. It is worth
mentioning that similar results are also obtained for progenitors of galaxies with a  mass of
$\geq 5\times 10^{11} {\rm M}_\odot$ at $z=0$.
Note again that the evolutions of $\gamma'_*$  estimating within [1-5] kpc or [5-10] kpc
are very similar.

\begin{figure}
\rotatebox{0}{\includegraphics[width=\columnwidth]{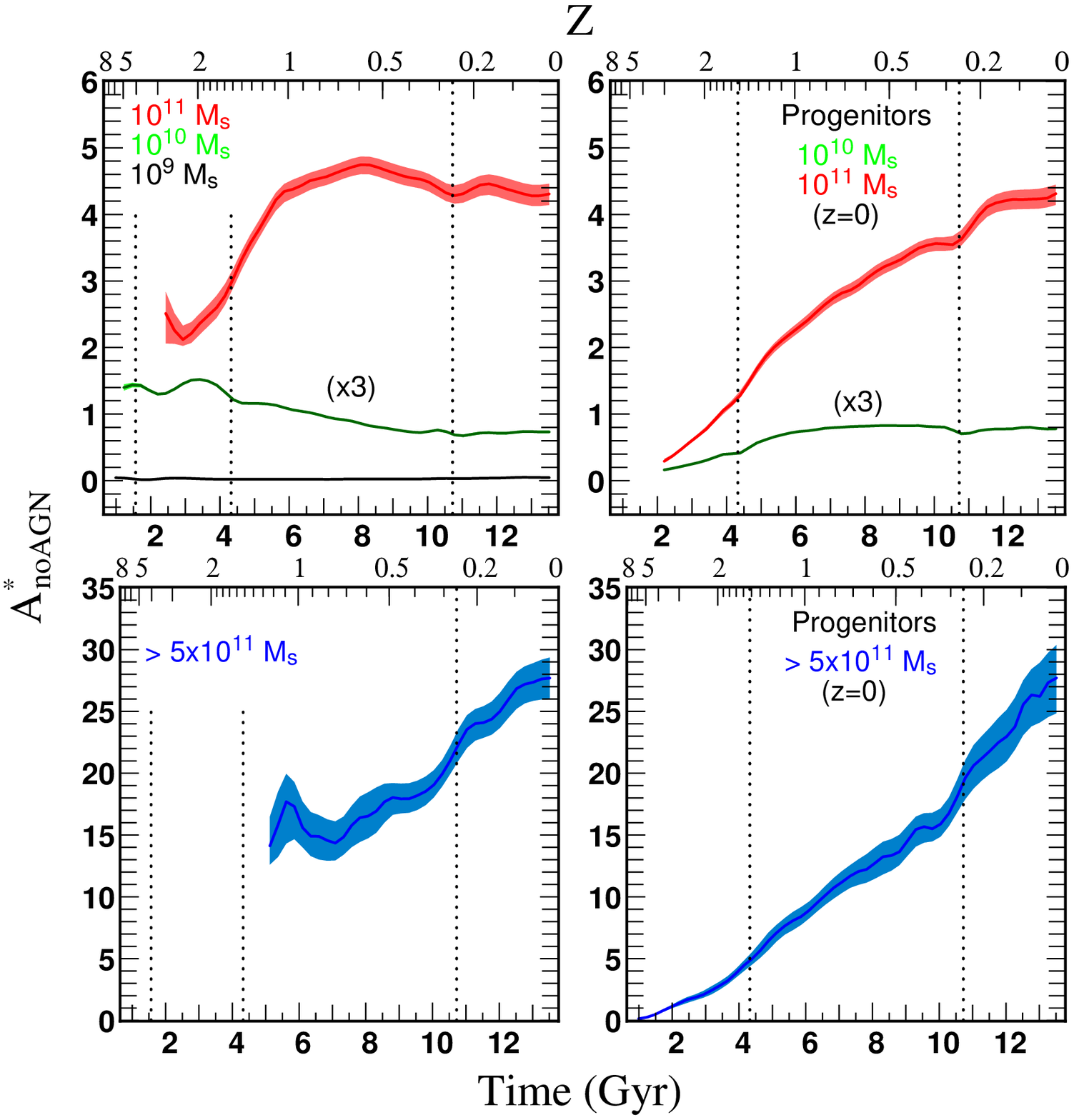}}
\caption{Time evolution of $A^*_{\rm noAGN}$ which measures the gap
between H$_{\rm AGN}$ and H$_{\rm noAGN}$ galaxy density profiles in the range of [1-5] kpc. 
The left panels consider galaxies within fixed mass intervals throughout cosmic time: 
$10^{9}$, $10^{10}$ and $10^{11} {\rm M}_\odot$  (upper panel) and
$\geq 5\times10^{11} {\rm M}_\odot $ (lower panel).
In the right panels, we follow the evolution of the
progenitors of galaxies that have a (H$_{\rm AGN}$) mass of $\sim 10^{10}$,
 $10^{11}$ and $\geq 5\times 10^{11} {\rm M}_\odot$ at $z=0$. Shaded
 areas represent the error
on the mean.
Vertical dashed lines
indicate when an additional level of refinement is introduced in the simulations. 
Note that we have multiplied results obtained for  $10^{10}  {\rm M}_\odot$ galaxies 
by a factor 3 for clarity. It is worth mentioning that similar evolutions
are obtained when estimating $A^*_{\rm noAGN}$ in the range of [1-10] kpc or [5-10]. Only the
amplitudes will of course change.}
\label{fig12}
 \end{figure}

Finally, Fig.  \ref{fig12}  shows the time evolution
of  $A^*_{\rm noAGN}$ for galaxies within a mass interval
fixed throughout cosmic time (right panels).
We use the same definition of $A_{\rm noAGN}$ (but for galaxies) given by equation \ref{equ2}
 and consider this time $r_1=1$ kpc and  $r_2=5$ kpc.
As expected,  for $10^{9} {\rm M}_\odot$ galaxies, $A_{\rm noAGN}$ stays
constant and equal to 0. On the contrary,
 for $10^{10} {\rm M}_\odot$ galaxies, $A^*_{\rm noAGN}$  is always positive and  
slightly decreasing below $z\sim 2$. Although AGN activity is
relatively weak in galaxies pertaining to this mass range,
it still affects their stellar density profiles all the way to the present time.
More massive galaxies, with masses of $10^{11} {\rm M}_\odot$ feature a rapid increase in
 $A^*_{\rm noAGN}$ until $z\sim 1$ and this quantity then remains roughly 
constant between $z=1$ and $z=0$. Finally,  $A^*_{\rm noAGN}$ is monotonously increasing for the most massive galaxies
($\geq 5\times10^{11} {\rm M}_\odot$). This is explained by two main
reasons. First, AGN activity rapidly flattens H$_{\rm AGN}$
galaxy density profiles at high redshift, and second, the mass of H$_{\rm noAGN}$ galaxies 
is still noticeably increasing down to $z=0$, as there is no upper limit in this  mass range.
Fig.  \ref{fig12} also presents the evolution of the progenitors of
these galaxies (left panels). In this case, $A^*_{\rm noAGN}$ is always
increasing (albeit more rapidly for more massive galaxies) which means that the gap between H$_{\rm AGN}$ and H$_{\rm noAGN}$ galaxy density profiles is continuously
increasing. 
Note again that the additional refinement levels 
have a limited but noticeable impact on the evolution of  $A^*_{\rm noAGN}$.
 The extra star formation spuriously induced at these epochs increases
 the central stellar mass, 
especially in  H$_{\rm noAGN}$ galaxies, which causes the more pronounced increases seen in
the evolution of  $A^*_{\rm noAGN}$. However these numerical effects
do not affect our conclusions.

\subsection{Matching or no matching, what is the difference?}

All along the paper so far, we have only carried out object-to-object comparisons
to ensure that we study the properties and evolution of the same objects in the different simulations. 
However, one can also be interested in comparing the evolution of
the properties of objects of similar mass between the three
simulations, which is not enforced by the matching strategy.
In this section we examine what the difference is between these two approaches.

\begin{figure}
\rotatebox{0}{\includegraphics[width=\columnwidth]{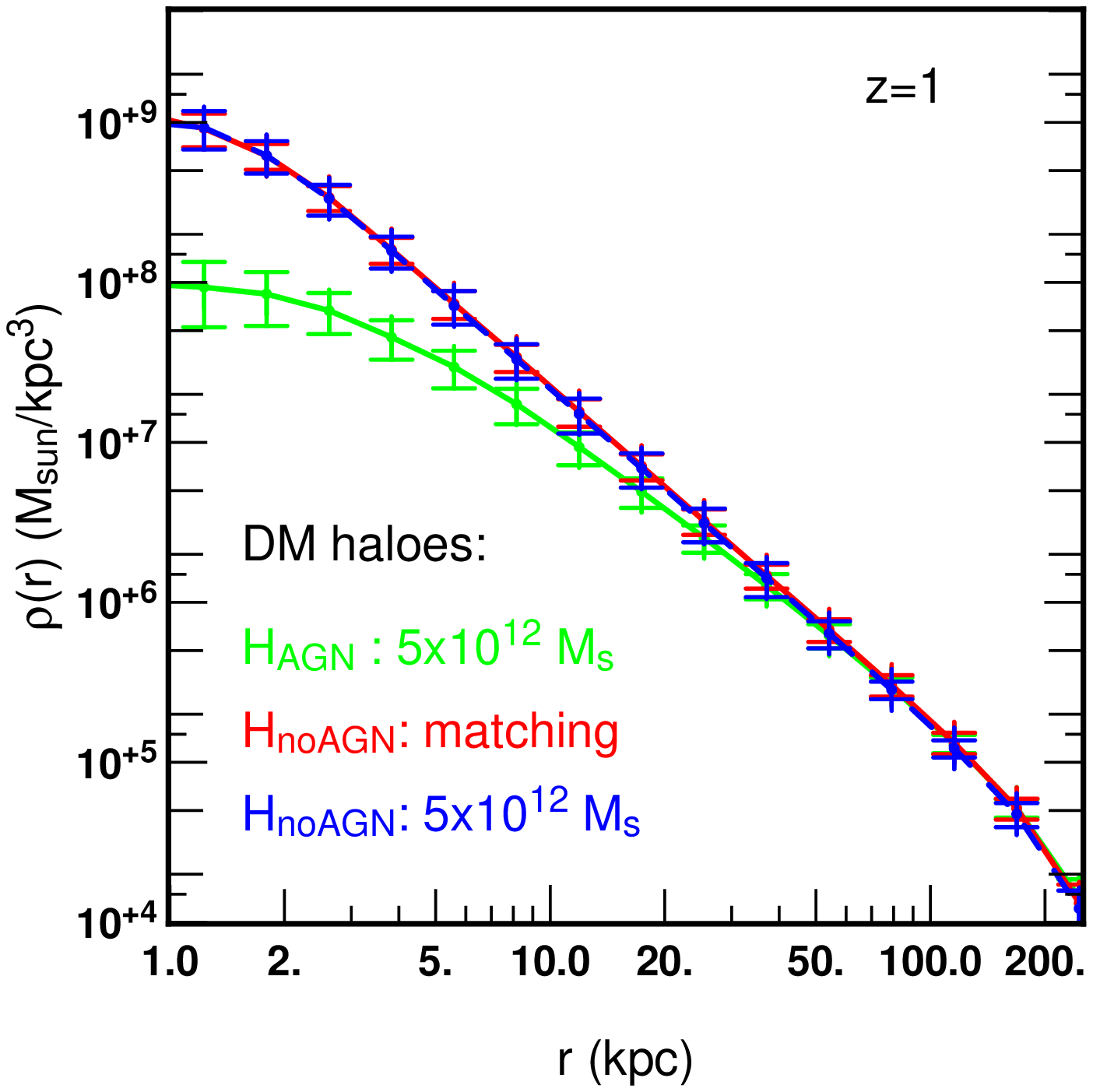}}
\rotatebox{0}{\includegraphics[width=\columnwidth]{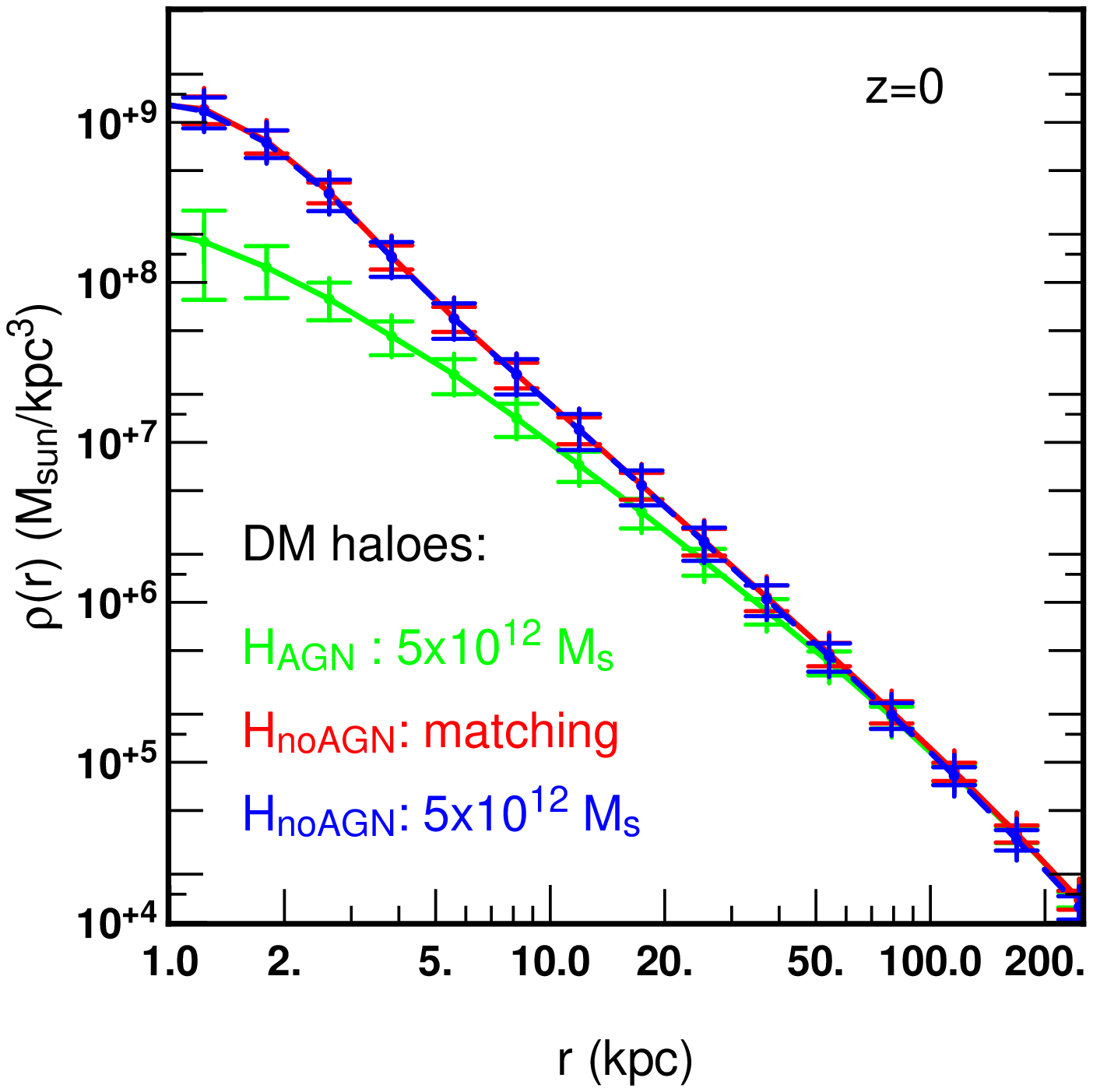}}
\caption{Mean density profiles of dark matter haloes with 
a mass of $5\times 10^{12} {\rm M}_\odot$  at $z=1$ (upper panel) and $z=0$ (lower panel) and
  extracted from the \hagnn (green)
or \hnoagnn (blue) simulations. The DM density profile of matched haloes is plotted in red. 
The error bars correspond to the standard deviations.
Since AGN feedback does not affect the virial mass of DM haloes, the red and blue profiles
sample the same population of haloes and are therefore indistinguishable.}
\label{fig13}
 \end{figure}

As far as the dark matter component is concerned, the answer is  
none,  because AGN feedback does not significantly affect the virial
mass of DM haloes. 
This is demonstrated in Fig. \ref{fig13} where we plot 
the mean density profiles of H$_{\rm noAGN}$ dark matter haloes either at $z=1$ or $z=0$,
either obtained through matching or simply considering the population
in the same mass interval: the two are indistinguishable from one another. 

\begin{figure}
\rotatebox{0}{\includegraphics[width=\columnwidth]{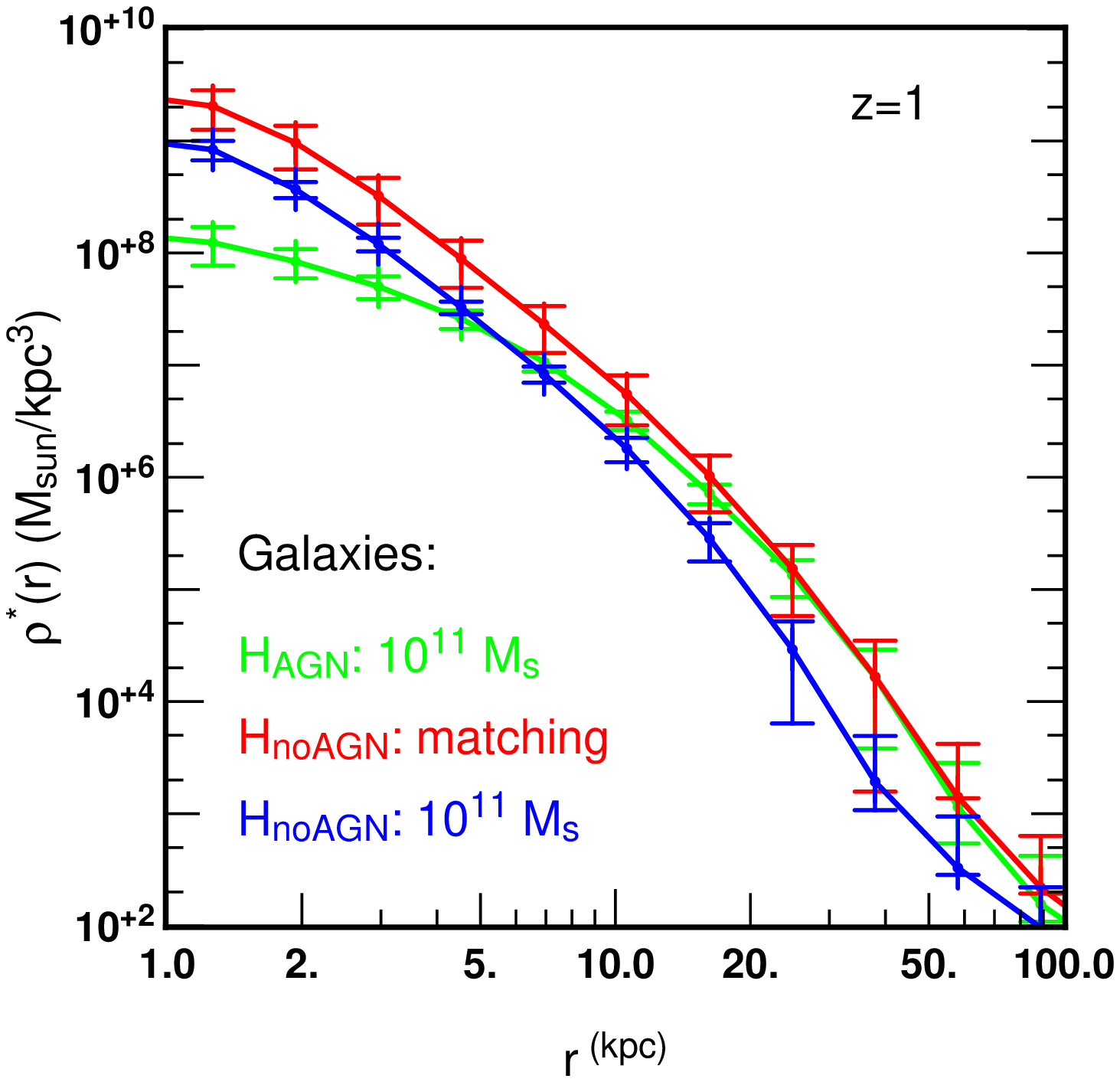}}
\rotatebox{0}{\includegraphics[width=\columnwidth]{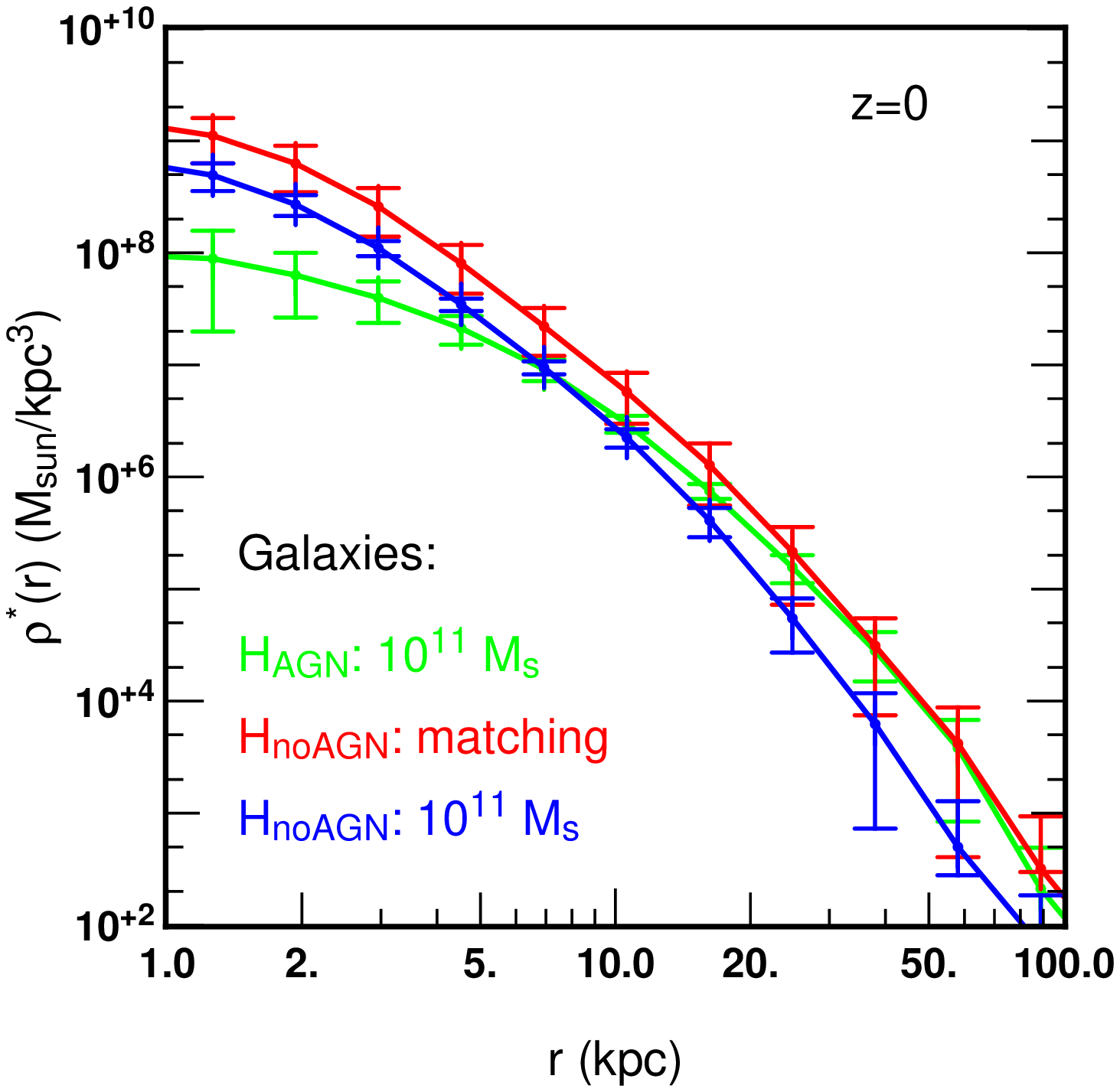}}
\caption{Mean stellar density profiles of galaxies with
a mass of $10^{11} {\rm M}_\odot$ at $z=1$ (upper panel) and $z=0$ (lower panel) and
 extracted from the \hagnn (green)
or \hnoagnn (blue) simulations. The stellar density profile of the
matched galaxies is plotted in red. The error bars  correspond to the 
standard deviations.
Contrary to the dark matter component, the presence of AGN feedback significantly
affects the virial mass of the galaxies. In this case, the red and blue profiles are not derived 
from the same population of galaxies and are therefore very different. }
\label{fig14}
 \end{figure}

On the contrary, stellar density profiles of galaxies prove to
 be very different. Indeed, for any given stellar mass interval of H$_{\rm AGN}$ galaxies,
the H$_{\rm noAGN}$ counterparts are in general much more massive. 
These galaxies display stellar density values that are much higher in the central parts
but similar at large radii to the  H$_{\rm AGN}$  ones.
Therefore, when comparing H$_{\rm AGN}$ and H$_{\rm noAGN}$ galaxies of the same mass,
Fig. \ref{fig14}  clearly shows that these latter still present 
higher central stellar density values. However, in order to compensate for
the extra mass enclosed within these regions, they also exhibit lower
densities at large radii. Thus, for a galaxy of a given mass,  
 AGN feedback leads to density profiles that are more centrally flat but also more extended.

\section{Discussion and Conclusions}

By comparing results from two state-of-the-art hydrodynamical
 cosmological simulations whose only difference is the
 presence/absence of AGN  feedback, and one cosmological simulation without baryons
which otherwise shares the same initial conditions,
we have explored the impact of AGN feedback on the
evolution of the inner density profiles of massive dark matter haloes and galaxies. 
We focused on dark matter haloes and galaxies with a mass greater than
$5\times 10^{11} {\rm M}_\odot$ and $10^{9} {\rm M}_\odot$ respectively. Since the resolution limit of the simulations
is 1 kpc (physical), we have only investigated the (relative) variations of 
halo and galaxy density profiles within a few kpc from their center (i.e. [1-5]kpc or [1-10]kpc).
Our findings can be summarized as follows:

$\bullet$  
When AGN feedback is included, the mean inner density profiles $\rho_{\rm AGN}$  of massive dark matter haloes 
 undergo successive phases of contraction (steepening) and expansion (flattening).
From very high redshift
to $z\sim 3$,  $\rho_{\rm AGN}$ becomes steeper than $\rho_{\rm DM}$ due to adiabatic contraction induced by 
early galaxy formation in the center of the host dark matter haloes.
 From $z\sim 3$ down  to $z\sim 1.6$, $\rho_{\rm AGN}$ is noticeably flattened 
by AGN activity which is high (quasar mode). From $z\sim 1.6$ to the present time, $\rho_{\rm AGN}$
steepens again  (``cusp regeneration'') as AGN activity considerably reduces.

$\bullet$ 
The gaps between  $\rho_{\rm AGN}$, $\rho_{\rm noAGN}$ and $\rho_{\rm DM}$ are also evolving  with time.
At high redshift,  $\rho_{\rm AGN}$ and $\rho_{\rm noAGN}$ tend to be much steeper than $\rho_{\rm DM}$ 
 due to rapid galaxy formation. Until $z\sim 1.6$, the flattening of $\rho_{\rm AGN}$ tends to
increase the gap with $\rho_{\rm noAGN}$  and conversely decrease the gap with $\rho_{\rm DM}$.
Finally after $z\sim 1.6$, the phase of ``cusp regeneration'' leads to  both a slight reduction and increase 
of the gap  w.r.t. $\rho_{\rm noAGN}$ and $\rho_{\rm DM}$ respectively.  

$\bullet$  
AGN feedback  noticeably reduces the central density in massive galaxies
and efficiently flattens their inner profiles, which lead to trends more compatible
with the observations of  massive elliptical
or cD galaxies, that exhibit very shallow
slopes in the stellar surface brightness profiles within small radii ($\approx 1$ kpc)
(Kormendy 1999; Quillen, Bower \& Stritzinger 2000; Laine et al. 2003; Graham 2004; Trujillo et al. 2004;
Lauer et al. 2005; Ferrarese et al. 2006; C\^ot\'e et al. 2007; Kormendy et al. 2009; Graham 2013).
In contrast  to the dark matter component, 
galaxy inner density profiles remain quite flat at low redshifts.

$\bullet$  
At any given redshift, more massive dark matter haloes or galaxies have in general
flatter central density profiles than their less massive counterparts. The impact of AGN feedback 
in the flattening of DM haloes/galaxies density profiles is all the
more important than the objects are massive.

$\bullet$  
Without AGN feedback, the inner density profiles of dark matter haloes and galaxies
are always very steep.

The present study clearly demonstrates that the inner density profiles of dark matter haloes and
galaxies are very sensitive to sub-grid physics and more specifically to AGN feedback.
For instance, our model predicts a 3-phase scenario in the evolution of dark matter density profiles
which is intimately  associated with the strength of AGN activity via the contribution
of the different accretion modes on the central BH. Indeed, as advocated by Peirani et al. (2008),
repetitive cycles of gas expansion by
AGN feedback and gas cooling are requested to efficiently flatten the dark matter profiles.
If the AGN activity progressively decreases, 
this proposed mechanism becomes less efficient at counterbalancing the DM adiabatic contraction 
and  at keeping the DM density profiles flat. Furthermore,  we  also found that
the radio mode tends to be the dominant mode after $z\sim 2$, which 
 renders the expansion phase of the gas more difficult.
In parallel, AGN feedback is also expected to regulate the star formation 
in the objects studied here (see Beckmann et al 2017 for detail).
 In this regard,  Kaviraj et al. (2017) have recently studied
the reproduction of quantities in \hagnn that trace the aggregate stellar-mass growth 
of galaxies over cosmic time namely the luminosity and stellar-mass functions,
 the star formation main sequence, rest-frame UV-optical-near infrared colours
 and the cosmic star-formation history. They found that \hagnn successfully
 captures the evolutionary trends of observed galaxies over
 the lifetime of the Universe, making it an excellent tool for studying the processes
  that drive galaxy evolution. However, it is worth mentioning that 
although the  galaxy stellar mass functions at different redshift
are consistent at the high mass end, \hagnn tend to overshoots the low-mass end (<$5\times10^{10} M_\odot$).
Similar conclusions were obtained by  Welker
et al. (2017) by comparing galaxy stellar mass functions with observation of CANDEL-UDS and GOODS surveys.
Future  accurate observations will allow to confirm these theoretical  predictions,
 but, above all, will help constraining AGN models.
One complication may arise when trying to probe the distribution of the
DM component. If one assumes that the
stellar mass distribution will trace that of the DM then high resolution observations
 of large samples of massive galaxies across a large
redshift range are requested. This could be done using EUCLID in the redshift range $z<1$ and
JWST in the redshift range $1<z<4$.

Basic comparisons with other theoretical works have also been considered in this work.
 For instance the mean density profiles of dark matter haloes derived from the Eagle 
simulation (Schayes et al. 2015) how noticeable differences especially for lower mass halos ($5\times10^{11} M_{\odot}$)
 but more consistent results for the massive ones. Moreover, we found that 
 the variations of the slope of the internal structure of DM haloes $\gamma'_{\rm dm}$ are also
in nice agreement with those derived from Eagle data in spite different hydrodynamic solvers
and AGN model implementations. 
Finally, our finding regarding the cusp regeneration phase
is in good agreement  results derived by Di Cintio et al. (2014) and
Tollet al. (2016) in hydrodynamical simulations
focusing on lower mass haloes (i.e. $M_{\rm halo}< 10^{12} {\rm M}_\odot$).
Further relevant comparison, 
with the Illustris simulation (Vogelsberger et al. 2014a; Genel et al. 2014) 
for instance,  will be done in a companion paper (Peirani et al. in prep)
focusing on the total density slope of massive early-type galaxies.

We mainly focused on the inner part of dark matter haloes density profiles
where clear differences can been seen between the different simulations.
Conversely, for a given mass sample, the density profiles of H$_{\rm AGN}$ haloes 
and their matching H$_{\rm noAGN}$ and H$_{\rm DM}$ haloes converge and become identical 
at distances $r>10-20$ kpc, which suggests that
these outer regions are not affected  by baryons and AGN feedback.
However, previous works indicate that
 AGN feedback is also expected to produce effect on larger scales.
For instance, AGN feedback is essential  to produce massive
galaxies that resemble ellipticals. Without BH activity, massive galaxies
 are disc-like with kinematics dominated by rotational support (see Dubois et al., 2013; Dubois et al. 2016).
At even larger scales, 
Suto et al. (2017) have recently  
examined the asphericity of galaxy clusters using the projected axis ratios of X-ray
 surface brightness, star, and dark matter distributions of the most massive haloes 
($M_{200}>5\times 10^{13} M_\odot$) extracted from \hagn, \hnoagnn and \hdm. 
They found that the baryonic physics and especially AGN feedback can significantly affect the
asphericity of dark matter distribution even beyond the central region, approximately up to
the half of the virial radius. 
AGN feedback seems therefore an indispensable ingredient for the formation of massive dark matter
haloes and galaxies.

In a companion paper, we will investigate the effect of AGN feedback on the total (DM + stars) density slope
of massive early-type galaxies estimated at the effective radius.

\vspace{1.0cm}

\noindent
{\bf Acknowledgements}

\noindent
We warmly thank the referee for an insightful
review that considerably improved the quality of the original manuscript.
We also warmly  thank Y.\,Suto and T.\,Nishimichi for interesting discussions
and M.\,Schaller for providing relevant Eagle simulation data.
S.\,P. acknowledges support from the Japan Society for the
 Promotion of Science (JSPS long-term invitation fellowship).
This  work  was  granted  access  to  the  HPC  resources  of
CINES under the allocations 2013047012, 2014047012 and
2015047012  made  by  GENCI and  has  made  use
of  the  Horizon  cluster hosted by the Institut d'Astrophysique de Paris
on  which  the  simulation  was post-processed.
 This work was carried out within the framework of the
Horizon project (\texttt{http://www.projet-horizon.fr})
and is partially supported by the grants ANR-13-BS05-0005 
of the French Agence Nationale de la Recherche.
The research of J.\,D. is supported by Adrian Beecroft and STFC.





\begin{thebibliography}{}



\bibitem[Abadi et al.(2010)]{2010MNRAS.407..435A}
Abadi, M.~G., Navarro, J.~F., Fardal, M., Babul, A., \& Steinmetz, M.\ 2010, \mnras, 407, 435 
\bibitem[Ahn \& Shapiro(2005)]{2005MNRAS.363.1092A}
Ahn, K., \& Shapiro, P.~R.\ 2005, \mnras, 363, 1092 
\bibitem[Aubert et al.(2004)]{aubert04} 
Aubert, D., Pichon, C., \& Colombi, S.\ 2004, \mnras, 352, 376 
\bibitem[Beckmann et al.(2017)]{2017arXiv170107838B}
Beckmann, R.~S., Devriendt, J., Slyz, A., et al.\ 2017, arXiv:1701.07838 
\bibitem[Blumenthal et al. 1986]{Blumenthaletal86}
Blumenthal, G.~R., Faber, S.~M., Flores, R., \& Primack, J.~R.\ 1986, \apj, 301, 27 
\bibitem[Bode et al. 2001]{bode01}
Bode, P., Ostriker, J.~P., \& Turok, N.\ 2001, \apj, 556, 93 
\bibitem[Bournaud et al.(2007)]{2007A&A...476.1179B}
Bournaud, F., Jog, C.~J., \& Combes, F.\ 2007, \aap, 476, 1179 
\bibitem[Bundy et al.(2015)]{2015ApJ...798....7B} 
Bundy, K., Bershady, M.~A., Law, D.~R., et al.\ 2015, \apj, 798, 7 
\bibitem[Burkert(2000)]{2000ApJ...534L.143B}
Burkert, A.\ 2000, \apjl, 534, L143 
\bibitem[Chan et al. 2015]{chanetal15}
Chan, T.~K., Kere{\v s}, D., O{\~n}orbe, J., et al.\ 2015, \mnras, 454, 2981 
\bibitem[Cheung et al.(2016)]{2016Natur.533..504C}
Cheung, E., Bundy, K., Cappellari, M., et al.\ 2016, \nat, 533, 504 
\bibitem[Chisari et al.(2017)]{2017arXiv170203913C}
Chisari, N.~E., Koukoufilippas, N., Jindal, A., et al.\ 2017, arXiv:1702.03913 
\bibitem[Cole et al. 2011]{coleetal11}
Cole, D.~R., Dehnen, W., \& Wilkinson, M.~I.\ 2011, \mnras, 416, 1118 
\bibitem[Col{\'{\i}}n et al. 2000]{colin00}
Col{\'{\i}}n, P., Avila-Reese, V., \& Valenzuela, O.\ 2000, \apj, 542, 622 
\bibitem[C{\^o}t{\'e} et al.(2007)]{2007ApJ...671.1456C} 
C{\^o}t{\'e}, P., Ferrarese, L., Jord{\'a}n, A., et al.\ 2007, \apj, 671, 1456 
\bibitem[Dav{\'e} et al.(2001)]{2001ApJ...547..574D}
Dav{\'e}, R., Spergel, D.~N., Steinhardt, P.~J., \& Wandelt, B.~D.\ 2001, \apj, 547, 574 
\bibitem[de Blok et al. 2001]{debloketal01}
de Blok W. J. G., McGaugh S. S., Bosma A. \& Rubin V. C., 2001, ApJ 552, L23
\bibitem[de Blok et al. 2008]{debloketal08}
de Blok, W.~J.~G., Walter, F., Brinks, E., et al.\ 2008, \aj, 136, 2648-2719 
\bibitem[Del Popolo(2009)]{2009ApJ...698.2093D}
Del Popolo, A.\ 2009, \apj, 698, 2093 
\bibitem[Del Popolo(2012)]{2012MNRAS.419..971D}
Del Popolo, A.\ 2012, \mnras, 419, 971 
\bibitem[Del Popolo(2014)]{2014JCAP...07..019D}
Del Popolo, A.\ 2014, \jcap, 7, 019 
\bibitem[Del Popolo \& Pace(2016)]{2016Ap&SS.361..162D}
Del Popolo, A., \& Pace, F.\ 2016, \apss, 361, 162 
\bibitem[Di Cintio et al.(2014)]{dicinto14}
Di Cintio, A., Brook, C.~B., Macci{\`o}, A.~V., et al.\ 2014, \mnras, 437, 415 
\bibitem[Dubois et al.(2010)]{2010MNRAS.409..985D}
Dubois, Y., Devriendt, J., Slyz, A., \& Teyssier, R.\ 2010, \mnras, 409, 985 
\bibitem[Dubois et al.(2012)]{duboisetal12}
Dubois, Y., Devriendt, J., Slyz, A., \& Teyssier, R.\ 2012, \mnras, 420, 2662 
\bibitem[Dubois et al.(2013)]{2013MNRAS.433.3297D}
Dubois, Y., Gavazzi, R., Peirani, S., \& Silk, J.\ 2013, \mnras, 433, 3297 
\bibitem[Dubois et al.(2014)]{duboisetal14}
Dubois, Y., Pichon, C., Welker, C., et al.\ 2014, \mnras, 444, 1453 
\bibitem[Dubois et al.(2016)]{2016MNRAS.463.3948D}
Dubois, Y., Peirani, S., Pichon, C., et al.\ 2016, \mnras, 463, 3948 
\bibitem[Duffy et al.(2010)]{2010MNRAS.405.2161D} 
Duffy, A.~R., Schaye, J., Kay, S.~T., et al.\ 2010, \mnras, 405, 2161 
\bibitem[Dutton \& Treu(2014)]{duttonettreu14}
Dutton, A.~A., \& Treu, T.\ 2014, \mnras, 438, 3594 
\bibitem[Elbert et al.(2015)]{2015MNRAS.453...29E} 
Elbert, O.~D., Bullock, J.~S., Garrison-Kimmel, S., et al.\ 2015, \mnras, 453, 29 
\bibitem[El-Zant et al. 2001]{el-zant1}
El-Zant, A., Shlosman, I., \& Hoffman, Y.\ 2001, \apj, 560, 636 
\bibitem[El-Zant et al. 2004]{el-zant2}
El-Zant, A.~A., Hoffman, Y., Primack, J., Combes, F., \& Shlosman, I.\ 2004, \apjl, 607, L75 
\bibitem[El-Zant et al. 2016]{el-zant3}
El-Zant, A.~A., Freundlich, J., \& Combes, F.\ 2016, \mnras, 461, 1745 
\bibitem[Faber et al.(1997)]{1997AJ....114.1771F}
Faber, S.~M., Tremaine, S., Ajhar, E.~A., et al.\ 1997, \aj, 114, 1771 
\bibitem[Ferrarese et al.(2006)]{2006ApJS..164..334F} 
Ferrarese, L., C{\^o}t{\'e}, P., Jord{\'a}n, A., et al.\ 2006, \apjs, 164, 334 
\bibitem[Genel et al.(2014)]{2014MNRAS.445..175G}
Genel, S., Vogelsberger, M., Springel, V., et al.\ 2014, \mnras, 445, 175 
\bibitem[Gentile et al. 2004]{gentileetal04}
Gentile, G., Salucci, P., Klein, U., Vergani, D., \& Kalberla, P.\ 2004, MNRAS, 351, 903 
\bibitem[Gnedin \& Zhao 2002]{gnedin}
Gnedin, O.~Y., \& Zhao, H.\ 2002, MNRAS, 333, 299 
\bibitem[Gnedin et al. 2004]{gnedin04}
Gnedin, O.~Y., Kravtsov, A.~V., Klypin, A.~A., \& Nagai, D.\ 2004, \apj, 616, 16 
\bibitem[Goerdt et al. 2006]{goerdtetal06}
Goerdt, T., Moore, B., Read, J.~I., Stadel, J., \& Zemp, M.\ 2006, \mnras, 368, 1073 
\bibitem[Governato et al. 2012]{governatoetal12}
Governato, F., Zolotov, A., Pontzen, A., et al.\ 2012, \mnras, 422, 1231 
\bibitem[Graham(2004)]{2004ApJ...613L..33G}
Graham, A.~W.\ 2004, \apjl, 613, L33 
\bibitem[Graham(2013)]{2013pss6.book...91G} 
Graham, A.~W.\ 2013, Planets, Stars and Stellar Systems.~Volume 6: Extragalactic Astronomy and Cosmology, 6, 91 
\bibitem[Gustafsson et al.(2006)]{2006PhRvD..74l3522G} 
Gustafsson, M., Fairbairn, M., \& Sommer-Larsen, J.\ 2006, \prd, 74, 123522 
\bibitem[Haardt \& Madau(1996)]{haardt&madau96}
Haardt, F., \& Madau, P.\ 1996, \apj, 461, 20 
\bibitem[Holley-Bockelmann et al. 2005]{holley-Bockelmann} 
Holley-Bockelmann, K., Weinberg, M., \& Katz, N.\ 2005, MNRAS, 363, 991 
\bibitem[Hopkins et al.(2007)]{2007ApJ...654..731H}
Hopkins, P.~F., Richards, G.~T., \& Hernquist, L.\ 2007, \apj, 654, 731 
\bibitem[Jardel \& Sellwood 2009]{jardel09}
Jardel, J.~R., \& Sellwood, J.~A.\ 2009, \apj, 691, 1300 
\bibitem[Jing \& Suto(2000)]{2000ApJ...529L..69J}
Jing, Y.~P., \& Suto, Y.\ 2000, \apjl, 529, L69 
\bibitem[Kaviraj et al.(2017)]{2017MNRAS.467.4739K}
Kaviraj, S., Laigle, C., Kimm, T., et al.\ 2017, \mnras, 467, 4739 
\bibitem[Khochfar \& Silk(2006)]{2006ApJ...648L..21K}
Khochfar, S., \& Silk, J.\ 2006, \apjl, 648, L21 
\bibitem[Kimm et al.(2012)]{2012MNRAS.425L..96K}
Kimm, T., Kaviraj, S., Devriendt, J.~E.~G., et al.\ 2012, \mnras, 425, L96 
\bibitem[Klypin et al.(2016)]{klypinetal16}
Klypin, A., Yepes, G., Gottl{\"o}ber, S., Prada, F., \& He{\ss}, S.\ 2016, \mnras, 457, 4340 
\bibitem[Kochanek \& White(2000)]{2000ApJ...543..514K}
Kochanek, C.~S., \& White, M.\ 2000, \apj, 543, 514 
\bibitem[Kormendy(1999)]{1999ASPC..182..124K}
Kormendy, J.\ 1999, Galaxy Dynamics - A Rutgers Symposium, 182,  
\bibitem[Kormendy et al.(2009)]{2009ApJS..182..216K}
Kormendy, J., Fisher, D.~B., Cornell, M.~E., \& Bender, R.\ 2009, \apjs, 182, 216 
\bibitem[Lackner \& Ostriker 2010]{lackner10}
Lackner, C.~N., \& Ostriker, J.~P.\ 2010, \apj, 712, 88 
\bibitem[Laine et al.(2003)]{2003AJ....125..478L} 
Laine, S., van der Marel, R.~P., Lauer, T.~R., et al.\ 2003, \aj, 125, 478 
\bibitem[Lauer et al.(2005)]{2005AJ....129.2138L}
Lauer, T.~R., Faber, S.~M., Gebhardt, K., et al.\ 2005, \aj, 129, 2138 
\bibitem[Leonard et al.(2007)]{2007ApJ...666...51L}
Leonard, A., Goldberg, D.~M., Haaga, J.~L., \& Massey, R.\ 2007, \apj, 666, 51 
\bibitem[Limousin et al.(2007)]{2007ApJ...668..643L}
Limousin, M., Richard, J., Jullo, E., et al.\ 2007, \apj, 668, 643 
\bibitem[Lin \& Loeb(2016)]{2016JCAP...03..009L}
Lin, H.~W., \& Loeb, A.\ 2016, \jcap, 3, 009 
\bibitem[Lovell et al.(2012)]{2012MNRAS.420.2318L}
Lovell, M.~R., Eke, V., Frenk, C.~S., et al.\ 2012, \mnras, 420, 2318 
\bibitem[Macci{\`o} et al. 2012]{maccioetal12}
Macci{\`o}, A.~V., Stinson, G., Brook, C.~B., et al.\ 2012, \apjl, 744, L9 
\bibitem[Macci{\`o} et al.(2015)]{2015MNRAS.453.1371M}
Macci{\`o}, A.~V., Mainini, R., Penzo, C., \& Bonometto, S.~A.\ 2015, \mnras, 453, 1371 
\bibitem[Marsh \& Pop (2015]{marshetpop15}
Marsh, D.~J.~E., \& Pop, A.-R.\ 2015, \mnras, 451, 2479 
\bibitem[Martizzi et al.(2012)]{2012MNRAS.422.3081M}
Martizzi, D., Teyssier, R., Moore, B., \& Wentz, T.\ 2012, \mnras, 422, 3081 
\bibitem[Martizzi et al.(2013)]{martizzi13}
Martizzi, D., Teyssier, R., \& Moore, B.\ 2013, \mnras, 432, 1947 
\bibitem[Mashchenko et al. 2006]{mashchenkoetal06}
Mashchenko, S., Couchman, H.~M.~P., \& Wadsley, J.\ 2006, \nat, 442, 539 
\bibitem[Mashchenko et al. 2008]{mashchenkoetal08}
Mashchenko, S., Wadsley, J., \& Couchman, H.~M.~P.\ 2008, Science, 319, 174 
\bibitem[Mazzalay et al.(2016)]{2016MNRAS.462.2847M}
Mazzalay, X., Thomas, J., Saglia, R.~P., et al.\ 2016, \mnras, 462, 2847 
\bibitem[Merritt 2004]{merrit}
Merritt, D. et al., \ 2004, ApJl, 607, L9
\bibitem[Merritt et al. 2006]{merrittetal06}
Merritt, D., Graham, A.~W., Moore, B., Diemand, J., \& Terzi{\'c}, B.\ 2006, \aj, 132, 2685 
\bibitem[Moore et al.(1998)]{1998ApJ...499L...5M}
Moore, B., Governato, F., Quinn, T., Stadel, J., \& Lake, G.\ 1998, \apjl, 499, L5 
\bibitem[Naab et al.(2009)]{2009ApJ...699L.178N}
Naab, T., Johansson, P.~H., \& Ostriker, J.~P.\ 2009, \apjl, 699, L178 
\bibitem[Navarro et al. 1996a]{nfw1}
Navarro, J.~F., Frenk, C.~S., \& White, S.~D.~M.\ 1996a, \apj, 462, 563 
\bibitem[Navarro et al. 1996b]{navarroetal96}
Navarro, J.~F., Eke, V.~R., \& Frenk, C.~S.\ 1996b, \mnras, 283, L72 
\bibitem[Navarro et al. 1997]{nfw2}
Navarro, J.~F., Frenk, C.~S., \& White, S.~D.~M.\ 1997, \apj, 490, 493 
\bibitem[Navarro et al. 2010]{navarroetal10}
Navarro, J.~F., Ludlow, A., Springel, V., et al.\ 2010, \mnras, 402, 21 
\bibitem[Newman et al.(2009)]{2009ApJ...706.1078N}
Newman, A.~B., Treu, T., Ellis, R.~S., et al.\ 2009, \apj, 706, 1078 
\bibitem[Newman et al.(2011)]{2011ApJ...728L..39N} 
Newman, A.~B., Treu, T., Ellis, R.~S., \& Sand, D.~J.\ 2011, \apjl, 728, L39 
\bibitem[Newman et al. 2013]{newmanetal13}
Newman, A.~B., Treu, T., Ellis, R.~S., \& Sand, D.~J.\ 2013, \apj, 765, 25 
\bibitem[Newman et al.(2015)]{2015ApJ...814...26N} 
Newman, A.~B., Ellis, R.~S., \& Treu, T.\ 2015, \apj, 814, 26 
\bibitem[Ogiya \& Mori 2011]{ogiyaetmori11}
Ogiya, G., \& Mori, M.\ 2011, \apjl, 736, L2 
\bibitem[Ogiya \& Mori 2014]{ogiyaetmori14}
Ogiya, G., \& Mori, M.\ 2014, \apj, 793, 46 
\bibitem[Oh et al. 2011]{ohetal11}
Oh, S.-H., de Blok, W.~J.~G., Brinks, E., Walter, F., \& Kennicutt, R.~C., Jr.\ 2011, \aj, 141, 193 
\bibitem[Oldham \& Auger(2016)]{2016MNRAS.457..421O}
Oldham, L.~J., \& Auger, M.~W.\ 2016, \mnras, 457, 421 
\bibitem[O{\~n}orbe et al. 2015]{onorbeetal15}
O{\~n}orbe, J., Boylan-Kolchin, M., Bullock, J.~S., et al.\ 2015, \mnras, 454, 2092 
\bibitem[Oser et al.(2010)]{2010ApJ...725.2312O}
Oser, L., Ostriker, J.~P., Naab, T., Johansson, P.~H., \& Burkert, A.\ 2010, \apj, 725, 2312 
\bibitem[Oser et al.(2012)]{2012ApJ...744...63O} 
Oser, L., Naab, T., Ostriker, J.~P., \& Johansson, P.~H.\ 2012, \apj, 744, 63 
\bibitem[Palunas \& Williams 2000]{palunasetwilliams00}
Palunas, P., \& Williams, T.~B.\ 2000, \aj, 120, 2884 
\bibitem[Pedrosa et al.(2010)]{2010MNRAS.402..776P}
Pedrosa, S., Tissera, P.~B., \& Scannapieco, C.\ 2010, \mnras, 402, 776 
\bibitem[Peirani et al.(2008)]{peirani08}
Peirani, S., Kay, S., \& Silk, J.\ 2008, \aap, 479, 123 
\bibitem[Peirani et al.(2010)]{2010MNRAS.405.2327P}
Peirani, S., Crockett, R.~M., Geen, S., et al.\ 2010, \mnras, 405, 2327 
\bibitem[Pontzen \& Governato 2012]{pontzen12}
Pontzen, A., \& Governato, F.\ 2012, \mnras, 421, 3464 
\bibitem[Power et al.(2003)]{power03} 
Power, C., Navarro, J.~F., Jenkins, A., et al.\ 2003, \mnras, 338, 14 
\bibitem[Quillen et al.(2000)]{2000ApJS..128...85Q}
Quillen, A.~C., Bower, G.~A., \& Stritzinger, M.\ 2000, \apjs, 128, 85 
\bibitem[Ragone-Figueroa \& Granato(2011)]{2011MNRAS.414.3690R}
 Ragone-Figueroa, C., \& Granato, G.~L.\ 2011, \mnras, 414, 3690 
\bibitem[Ragone-Figueroa et al.(2012)]{2012MNRAS.423.3243R}
Ragone-Figueroa, C., Granato, G.~L., \& Abadi, M.~G.\ 2012, \mnras, 423, 3243 
\bibitem[Ragone-Figueroa et al.(2013)]{2013MNRAS.436.1750R} 
Ragone-Figueroa, C., Granato, G.~L., Murante, G., Borgani, S., \& Cui, W.\ 2013, \mnras, 436, 1750 
\bibitem[Read \& Gilmore 2005]{readetgilmore05}
Read, J.~I., \& Gilmore, G.\ 2005, \mnras, 356, 107 
\bibitem[Richtler et al.(2011)]{2011A&A...531A.119R}
Richtler, T., Salinas, R., Misgeld, I., et al.\ 2011, \aap, 531, A119 
\bibitem[Rodriguez-Gomez et al.(2016)]{2016MNRAS.458.2371R}
Rodriguez-Gomez, V., Pillepich, A., Sales, L.~V., et al.\ 2016, \mnras, 458, 2371 
\bibitem[Romano-D{\'{\i}}az et al.(2008)]{2008ApJ...685L.105R}
Romano-D{\'{\i}}az, E., Shlosman, I., Hoffman, Y., \& Heller, C.\ 2008, \apjl, 685, L105 
\bibitem[Salucci \& Burkert 2000]{salucciburket00}
Salucci, P., \& Burkert, A.\ 2000, ApJl, 537, L9 
\bibitem[Sand et al. 2004]{sandetal04}
Sand, D.~J., Treu, T., Smith, G.~P., \& Ellis, R.~S.\ 2004, \apj, 604, 88 
\bibitem[Sand et al.(2008)]{2008ApJ...674..711S}
Sand, D.~J., Treu, T., Ellis, R.~S., Smith, G.~P., \& Kneib, J.-P.\ 2008, \apj, 674, 711-727 
\bibitem[Schaller et al.(2015)]{2015MNRAS.451.1247S} 
Schaller, M., Frenk, C.~S., Bower, R.~G., et al.\ 2015a, \mnras, 451, 1247    %
\bibitem[Schaller et al.(2015)]{2015MNRAS.452..343S}
 Schaller, M., Frenk, C.~S., Bower, R.~G., et al.\ 2015b, \mnras, 452, 343    
\bibitem[Schaye et al.(2015)]{2015MNRAS.446..521S}
 Schaye, J., Crain, R.~A., Bower, R.~G., et al.\ 2015, \mnras, 446, 521 
\bibitem[Sellwood 2008]{sellwood}
Sellwood, J.~A.\ 2008, \apj, 679, 379-396 
\bibitem[Shankar et al.(2009)]{2009ApJ...690...20S}
Shankar, F., Weinberg, D.~H., \& Miralda-Escud{\'e}, J.\ 2009, \apj, 690, 20 
\bibitem[Shankar et al.(2013)]{2013MNRAS.428..109S}
Shankar, F., Marulli, F., Bernardi, M., et al.\ 2013, \mnras, 428, 109 
\bibitem[Spekkens et al. 2005]{spekkensetal05}
Spekkens, K., Giovanelli, R., \& Haynes, M.~P.\ 2005, \aj, 129, 2119 
\bibitem[Spergel \& Steinhardt 2000]{spergel00}
Spergel, D.~N., \& Steinhardt, P.~J.\ 2000, Physical Review Letters, 84, 3760 
\bibitem[Stadel et al. 2009]{stadeletal09}
Stadel, J., Potter, D., Moore, B., et al.\ 2009, \mnras, 398, L21 
\bibitem[Sutherland \& Dopita (1993)]{sutherland&dopita93}
Sutherland, R.~S., \& Dopita, M.~A.\ 1993, \apjs, 88, 253 
\bibitem[Suto et al.(2017)]{2017PASJ...69...14S}
Suto, D., Peirani, S., Dubois, Y., et al.\ 2017, \pasj, 69, 14 
\bibitem[Swaters et al. 2003]{swatersetal03}
Swaters, R.~A., Madore, B.~F., van den Bosch, F.~C., \& Balcells, M.\ 2003, \apj, 583, 732 
\bibitem[Teyssier(2002)]{2002A&A...385..337T}
Teyssier, R.\ 2002, \aap, 385, 337 
\bibitem[Teyssier et al.(2011)]{2011MNRAS.414..195T}
Teyssier, R., Moore, B., Martizzi, D., Dubois, Y., \& Mayer, L.\ 2011, \mnras, 414, 195 
\bibitem[Teyssier et al.(2013)]{2013MNRAS.429.3068T}
Teyssier, R., Pontzen, A., Dubois, Y., \& Read, J.~I.\ 2013, \mnras, 429, 3068 
\bibitem[Thomas et al.(2014)]{2014ApJ...782...39T} 
Thomas, J., Saglia, R.~P., Bender, R., Erwin, P., \& Fabricius, M.\ 2014, \apj, 782, 39 
\bibitem[Tissera et al.(2010)]{2010MNRAS.406..922T}
Tissera, P.~B., White, S.~D.~M., Pedrosa, S., \& Scannapieco, C.\ 2010, \mnras, 406, 922 
\bibitem[Tollet et al.(2016)]{tolletetal16}
Tollet, E., Macci{\`o}, A.~V., Dutton, A.~A., et al.\ 2016, \mnras, 456, 3542 
\bibitem[Tonini et al. 2006]{tonini}
Tonini, C., Lapi, A., \& Salucci, P.\ 2006, ApJ, 649, 591 
\bibitem[Trujillo et al.(2004)]{2004AJ....127.1917T}
Trujillo, I., Erwin, P., Asensio Ramos, A., \& Graham, A.~W.\ 2004, \aj, 127, 1917 
\bibitem[Tweed et al.(2009)]{tweed09} 
Tweed, D., Devriendt, J., Blaizot, J., Colombi, S., \& Slyz, A.\ 2009, \aap, 506, 647 
\bibitem[Ueda et al.(2014)]{2014ApJ...786..104U}
Ueda, Y., Akiyama, M., Hasinger, G., Miyaji, T., \& Watson, M.~G.\ 2014, \apj, 786, 104 
\bibitem[Umetsu et al.(2007)]{2007MPLA...22.2099U} 
Umetsu, K., Takada, M., \& Broadhurst, T.\ 2007, Modern Physics Letters A, 22, 2099 
\bibitem[Vogelsberger et al.(2014)]{2014MNRAS.444.1518V}
 Vogelsberger, M., Genel, S., Springel, V., et al.\ 2014a, \mnras, 444, 1518 
\bibitem[Vogelsberger et al. 2014]{vogelsbergeretal14}
Vogelsberger, M., Zavala, J., Simpson, C., \& Jenkins, A.\ 2014b, \mnras, 444, 3684 
\bibitem[Volonteri et al.(2016)]{2016MNRAS.460.2979V}
Volonteri, M., Dubois, Y., Pichon, C., \& Devriendt, J.\ 2016, \mnras, 460, 2979 
\bibitem[Walter et al. 2008]{walteretal08}
Walter, F., Brinks, E., de Blok, W.~J.~G., et al.\ 2008, \aj, 136, 2563-2647 
\bibitem[Walker \& Pe{\~n}arrubia 2011]{walkeretpenarrubia11}
Walker, M.~G., \& Pe{\~n}arrubia, J.\ 2011, \apj, 742, 20
\bibitem[Weinberg \& Katz 2002]{weinberg}
Weinberg, M.~D., \& Katz, N.\ 2002, ApJ, 580, 627 
\bibitem[Weinberger et al.(2017)]{2017MNRAS.465.3291W}
Weinberger, R., Springel, V., Hernquist, L., et al.\ 2017, \mnras, 465, 3291 
\bibitem[Welker et al.(2017)]{2017MNRAS.465.1241W}
Welker, C., Dubois, Y., Devriendt, J., et al.\ 2017, \mnras, 465, 1241 
\bibitem[Yoshida et al.(2000)]{2000ApJ...535L.103Y}
Yoshida, N., Springel, V., White, S.~D.~M., \& Tormen, G.\ 2000, \apjl, 535, L103 
\bibitem[Zitrin et al.(2015)]{2015ApJ...801...44Z}
Zitrin, A., Fabris, A., Merten, J., et al.\ 2015, \apj, 801, 44 







 








\end{thebibliography}
\end{document}